\documentclass[11pt]{amsart}

\usepackage[margin=2.8cm]{geometry}
\usepackage{amsmath,amssymb,amsthm,mathtools}
\usepackage{enumitem}
\usepackage{graphicx}
\usepackage{booktabs}
\usepackage[colorlinks=true,linkcolor=blue,citecolor=blue,urlcolor=blue]{hyperref}
\usepackage[section]{placeins}
\theoremstyle{plain}
\newtheorem{theorem}{Theorem}[section]
\newtheorem{proposition}[theorem]{Proposition}
\newtheorem{lemma}[theorem]{Lemma}
\newtheorem{corollary}[theorem]{Corollary}
\theoremstyle{definition}
\newtheorem{definition}[theorem]{Definition}
\newtheorem{example}[theorem]{Example}
\newtheorem{remark}[theorem]{Remark}
\numberwithin{equation}{section}

\newcommand{\R}{\mathbb{R}}
\newcommand{\sym}{\mathrm{sym}}
\newcommand{\skw}{\mathrm{sk}}
\DeclareMathOperator{\diag}{diag}
\DeclareMathOperator{\tr}{tr}
\DeclareMathOperator{\dv}{div}
\DeclareMathOperator{\Cov}{Cov}
\DeclareMathOperator{\Var}{Var}

\begin{document}

\title[Metriplectic Lotka--Volterra systems with noise]{Metriplectic Lotka--Volterra systems with noise:\\ interactions invisible at stationarity, visible in time}

\author{Jose Á. ~Capit\'an}
\address{Department of Applied Mathematics, E.T.S.\ de Edificaci\'on, Universidad Polit\'ecnica de Madrid, Avenida Juan de Herrera 6, 28040 Madrid, Spain}
\email{ja.capitan@upm.es}

\author{Alberto ~Meg\'ias}
\address{Department of Applied Mathematics, E.T.S.\ de Edificaci\'on, Universidad Polit\'ecnica de Madrid, Avenida Juan de Herrera 6, 28040 Madrid, Spain}
\email{alberto.megias@upm.es}

\author{Cristina ~Sard\'on}
\address{Department of Applied Mathematics, E.T.S.\ de Edificaci\'on, Universidad Polit\'ecnica de Madrid, Avenida Juan de Herrera 6, 28040 Madrid, Spain}
\email{mariacristina.sardon@upm.es}

\date{\today}

\begin{abstract}
We show that a Lotka--Volterra system with an interior equilibrium admits a metriplectic structure, with a Volterra function $V_C$ as energy and an entropy variable for the environment, if and only if Volterra's dissipativity condition $CA+A^{\top}C\le0$ holds, and we construct it explicitly with a log-canonical Poisson bivector. Adding noise through the fluctuation--dissipation relation of stochastic GENERIC yields multiplicative environmental noise whose covariance is fixed by the symmetric part of $AC^{-1}$, and whose stationary law is an explicit Gibbs measure, a product of independent Gamma distributions. Unlike the Gibbs measures known for stochastic Lotka--Volterra systems with symmetric interactions, this holds for nonsymmetric interactions, predator--prey interactions included. We then show that the metriplectic decomposition coincides with a decomposition of what can be observed. The stationary distribution determines only the equilibrium and the matrix $C/k$: every interaction matrix in an open cone of dimension $n^2$, and in particular any reversible (Poisson) part of the interaction, produces (with its compatible noise) exactly the same stationary law, which is also the law of a system of uncoupled stochastic logistic equations. Gamma-distributed and uncorrelated abundance fluctuations, as observed in microbial communities, are therefore no formal evidence of the absence of interactions. The interactions are, however, fully visible in time: the derivative at zero lag of the covariance of the log-abundances equals $A(C/k)^{-1}$, its antisymmetric part is the Poisson part of the structure, and the stationary entropy production rate, which we compute in closed form, is carried entirely by the Poisson part. Finally, for some noise we prove an exact Lyapunov-type identity $AK+K^\top A^\top+Q=0$ for the cross-covariance $K=\Cov(x,\log x)$ of any stationary solution. It implies that the stationary populations are independent if and only if the noise is compatible with a metriplectic structure, and it gives an explicit lower bound for the cross-covariances between species in terms of the distance of the noise covariance $Q$ to the compatible family. Simulations of a predator--prey model confirm these results and show that the bound is nearly sharp.
\end{abstract}

\maketitle

\section{Introduction}

Metriplectic systems \cite{Kau84,Mor84,Mor86}, also known as GENERIC (general equation for the non-equilibrium reversible and irreversible coupling) systems \cite{GO97,Ott05}, couple a reversible dynamics, generated by a Poisson bivector and an energy function, with an irreversible one, generated by a symmetric tensor and an entropy function, in such a way that the energy is conserved and the entropy does not decrease. They admit a natural stochastic extension, in which the noise and the metric part are tied together by a fluctuation--dissipation relation \cite{Ott05,PS25}. In this paper we apply this framework to Lotka--Volterra (LV) systems, and we show that it answers a concrete question of community ecology: what can be learned about the interactions between species from fluctuating abundance data.

LV systems have a well-known Poisson structure in the conservative case \cite{Vol31,Pla95}, and Volterra-type Lyapunov functions in the dissipative case \cite{Vol31,HS98,Tak96}; for stably dissipative systems, the dynamics on the attractor is Hamiltonian with respect to a log-canonical Poisson structure \cite{DFO98}. On the statistical side, Kerner \cite{Ker57,Ker59} constructed a statistical mechanics of conservative Volterra systems, using Volterra's integral as an energy and the Liouville measure in logarithmic variables; see also \cite{GMM71}. Metriplectic formulations of particular LV models have been considered in the literature \cite{IIP11}, and stochastic LV models with multiplicative environmental noise have been studied extensively, including sharp criteria for persistence and extinction \cite{MSR03,HN18}. In the statistical physics of large disordered ecosystems \cite{Bun17,BBC18,ARCB21}, stochastic LV systems with demographic noise and \emph{symmetric} interactions are known to have a Boltzmann-type stationary distribution, which makes them amenable to the methods of equilibrium statistical mechanics and spin glasses \cite{ARCB21,GH25}. For nonsymmetric interactions, which are the rule in real communities, as in predator--prey pairs, no such measure is available in general.

Our ecological motivation is the following. In microbial communities, the distribution of abundance fluctuations of each species is well described by a Gamma distribution, and the stochastic logistic model, in which species do not interact and are driven by environmental noise, reproduces this and other macroecological patterns \cite{Gri20}. The success of a model without interactions in communities where interactions are known to exist calls for an explanation. On the other hand, interactions are routinely estimated from time series of log-abundances by multivariate autoregressive (MAR) models \cite{Ives03}, which are linear approximations around equilibrium. The results of this paper explain the first observation and give an exact, nonlinear foundation for the second: stationary statistics, which describe the community at a single time, are blind to a large part of the interactions, whereas the time ordering of the fluctuations reveals them completely.

\subsection*{Main results}
We list our contributions, indicating in each case how they relate to the existing literature.
\begin{enumerate}[label=(\roman*)]
\item \emph{Classification} (Section~\ref{sec:LV}). An LV system with an interior equilibrium admits a metriplectic structure with energy $V_C+s$ and entropy $s$, where $V_C$ is a Volterra function with positive diagonal weights $C$ and $s$ is an entropy variable for the environment, if and only if Volterra's dissipativity condition $CA+A^\top C\le0$ holds. Sufficiency is proved by an explicit construction, in which the log-canonical bivector is given by the skew-symmetric part of $AC^{-1}$ and the metric by its symmetric part (Theorem~\ref{thm:metriplectic}): the dynamics splits into a reversible cycle, the conservative LV system, and a friction, intraspecific competition. Necessity holds for an arbitrary bivector and an arbitrary metric (Proposition~\ref{prop:necessity}). The ingredients are classical \cite{Vol31,Pla95,DFO98}; what is new is the metriplectic formulation of general dissipative LV systems, and the fact that no choice of bivector or metric can avoid Volterra's condition.

\item \emph{Selection of the noise} (Section~\ref{sec:noise}). Stochastic LV models usually prescribe the environmental noise independently of the deterministic dynamics. The fluctuation--dissipation relation of stochastic GENERIC instead selects a family of \emph{compatible} noises, whose covariance is fixed by $A$, $C$ and a temperature $k$ (Corollary~\ref{cor:FDT} and Remark~\ref{rem:family}): the noise has the standard multiplicative form used in population dynamics, but the noise acting on each species is proportional to its self-limitation, $\sigma_i^2=2k|A_{ii}|/c_i$. With this noise the energy $V_C+s$ is conserved exactly (Theorem~\ref{thm:stochastic}). The construction requires a choice of reference measure, which we discuss in Remark~\ref{rem:measure}.

\item \emph{An explicit Gibbs measure for nonsymmetric interactions} (Theorem~\ref{thm:stochastic}). With compatible noise, the stationary law is a product of independent Gamma distributions, for every interaction matrix satisfying Volterra's condition, predator--prey interactions included. In the statistical physics of disordered ecosystems, Boltzmann-type stationary measures are available only for symmetric interactions \cite{BBC18,ARCB21,GH25}; Theorem~\ref{thm:stochastic} provides an explicit Gibbs measure for a class of nonsymmetric ones, through a different mechanism based on environmental noise (Remark~\ref{rem:disordered}).

\item \emph{Stationary blindness} (Section~\ref{sec:blind}). Given a product of Gamma distributions, the interaction matrices that produce it as the stationary law of a compatible system form an open cone of dimension $n^2$, in which the reversible part of the interaction is completely free (Theorem~\ref{thm:blind}). In particular, every such system has exactly the same stationary law as a system of uncoupled stochastic logistic equations driven by the same noise (Corollary~\ref{cor:twin}): a predator--prey pair and two species that do not interact at all cannot be distinguished from their abundance distributions. This explains why a model without interactions can reproduce the Gamma statistics of microbial abundances \cite{Gri20}, and shows that the agreement does not rule out interactions (Remark~\ref{rem:SLM}). The proof is short once Theorem~\ref{thm:stochastic} is available; This provides a novel perspective on the interpretation of ecological data

\item \emph{Dynamical identifiability} (Section~\ref{sec:time}). For any stationary stochastic LV system with constant noise covariance, the derivative at zero lag of the covariance $R(\tau)$ of the log-abundances satisfies the exact identity $\dot R(0^+)=A\Cov(x,\log x)$ (Theorem~\ref{thm:moments}), so that the interaction matrix is recovered from one-time moments and short-lag correlations (Corollary~\ref{cor:identify}). The idea of estimating interactions from time series is that of MAR models \cite{Ives03}, which rely on a linearization around equilibrium; Theorem~\ref{thm:moments} is an exact, nonlinear, continuous-time counterpart for LV systems, valid for any noise intensity (Remark~\ref{rem:MAR}). For compatible noise, the reversible part of the interaction, which is invisible at stationarity, is exactly the time asymmetry of the lagged covariances. In our simulations the interaction matrix is recovered with errors of a few percent.

\item \emph{The arrow of time} (Theorem~\ref{thm:ep}). The stationary entropy production rate of the compatible system, which measures how far the stationary process is from being time-reversible, is given in closed form, does not depend on $k$, and vanishes if and only if the Poisson part of the structure vanishes. Thus, while in the deterministic system the entropy is produced by the metric part, the whole irreversibility of the stationary stochastic process is carried by the Poisson part (Remark~\ref{rem:duality}).The computation is straightforward, but the explicit formula and this reversal of roles do not appear to have been previously documented.

\item \emph{Rigidity and incompatible noise} (Section~\ref{sec:incompatible}). For an arbitrary constant noise covariance $Q$, the cross-covariance $K=\Cov(x,\log x)$ of any stationary solution satisfies the exact Lyapunov-type identity $AK+K^\top A^\top+Q=0$ (Theorem~\ref{thm:exact}). Two consequences follow. First, the stationary populations are independent if and only if the noise is compatible (Corollary~\ref{cor:independence}), so that the product form of the stationary law characterizes the stochastic metriplectic systems exactly. Second, the cross-covariances between species are bounded below by the distance of $Q$ to the compatible family (Proposition~\ref{prop:bound}), and the bound is nearly sharp in our simulations. In the linear-noise approximation the sign and size of the correlations are also given explicitly (Theorem~\ref{thm:linear} and Corollary~\ref{cor:pp-corr}). Stationary correlations between species are therefore a signature of the mismatch between noise and dissipation, rather than of interaction. This is the part of the paper with the most substantial mathematical content of its own.
\end{enumerate}

Several of the computations below are instances of general facts, and we wish to be explicit about this. In the logarithmic coordinates $u=\log x$, the population part of the compatible stochastic system is a Langevin equation with a constant nonreversible perturbation, of the type studied in \cite{HHS93,RBS15}, and Theorem~\ref{thm:moments} is an elementary consequence of It\^o's formula. What the metriplectic formulation adds is the identification of the relevant decomposition: it selects the coordinates, the noise and the reference measure, and it matches the three geometric ingredients of the structure with three observable features of the fluctuations, namely the Volterra function with the stationary law, the metric with the quadratic variation of the noise, and the Poisson bivector with the time asymmetry. Results (iii), (iv), (vi), and (vii) follow from this correspondence and provide further consequences of this relationship.

We stress that compatible noise is used here as an exactly solvable reference case, not as a claim about real environmental noise. For a single species every environmental noise is compatible, but for $n\ge2$ compatibility ties the noise intensities to self-limitation, and generic noise does not satisfy it. Section~\ref{sec:incompatible} quantifies the departure from the reference case in terms of observable correlations, and Corollary~\ref{cor:independence} turns compatibility itself into a testable hypothesis.

The paper is organized as follows. Section~\ref{sec:metriplectic} recalls metriplectic systems. Section~\ref{sec:LV} constructs the metriplectic structure of LV systems, and Section~\ref{sec:noise} introduces the compatible noise and its stationary law. Sections~\ref{sec:blind} and~\ref{sec:time} contain the results on stationary blindness and dynamical identifiability, and Section~\ref{sec:incompatible} treats incompatible noise. Section~\ref{sec:numerics} illustrates the results numerically on a predator--prey model, and the last section contains concluding remarks.

\section{Metriplectic systems}\label{sec:metriplectic}

\begin{definition}\label{def:metriplectic}
A \emph{metriplectic system} on a manifold $Q$ is a quadruple $(\Pi,M,E,S)$, where $\Pi$ is a Poisson bivector on $Q$, $M$ is a symmetric positive semidefinite bilinear form on $T^*Q$, and $E,S\in C^\infty(Q)$, such that
\begin{equation}\label{eq:degeneracy}
\Pi^\sharp(dS)=0,\qquad M^\sharp(dE)=0 .
\end{equation}
Its dynamics is given by the vector field
\begin{equation}\label{eq:flow}
\dot z=\Pi^\sharp(dE)+M^\sharp(dS),\qquad z\in Q .
\end{equation}
\end{definition}

Here $z$ denotes a point of $Q$, and in local coordinates $\Pi^\sharp(dF)^a=\Pi^{ab}\partial_bF$ and $M^\sharp(dF)^a=M^{ab}\partial_bF$. Equivalently, if $\mathcal F=E+S$, then $\dot F=\{F,\mathcal F\}+(F,\mathcal F)$ for every function $F$, where $\{F,G\}=\Pi(dF,dG)$ and $(F,G)=M(dF,dG)$; by \eqref{eq:degeneracy}, the Poisson bracket only sees $E$ and the symmetric bracket only sees $S$.

Each condition in Definition~\ref{def:metriplectic} has a clear meaning. Skew-symmetry of $\Pi$ implies that the reversible part conserves $E$, and the Jacobi identity guarantees that this part is a genuine Hamiltonian flow, so that its flows preserve the bivector, its Casimir functions and, when $\Pi$ is unimodular, a Liouville measure; the last property will be essential in Section~\ref{sec:noise}. The condition $\Pi^\sharp(dS)=0$ says that $S$ is a Casimir function, so that the reversible part produces no entropy, and $M^\sharp(dE)=0$ says that the irreversible part does not change the total energy. Finally, positivity of $M$ gives the second law.

\begin{lemma}\label{lem:laws}
Along the solutions of \eqref{eq:flow}, $\dot E=0$ and $\dot S=M(dS,dS)\ge0$.
\end{lemma}

\begin{proof}
We have $\dot E=\Pi(dE,dE)+M(dE,dS)$. The first term vanishes by skew-symmetry and the second one equals $\langle dS,M^\sharp dE\rangle=0$. Similarly, $\dot S=\Pi(dS,dE)+M(dS,dS)=-\langle dE,\Pi^\sharp dS\rangle+M(dS,dS)=M(dS,dS)$.
\end{proof}

\begin{remark}\label{rem:dimension}
A metriplectic system with $\Pi\not\equiv0$ and $dS\neq0$ on a dense subset of $Q$ requires $\dim Q\ge3$. Indeed, $S$ must be a Casimir function of $\Pi$. In dimension one every bivector vanishes. In dimension two we can write locally $\Pi=f\,\partial_1\wedge\partial_2$, and $\Pi^\sharp(dS)=0$ gives $f\,dS=0$, so $f$ vanishes on the dense set where $dS\neq0$, and hence everywhere by continuity. (Without the density assumption the statement fails in the smooth category, since $\Pi$ could be supported in the interior of a region where $S$ is constant.) In particular, for $S=s$, as in the construction below, $dS$ never vanishes and $\Pi\equiv0$. Thus a single population, even after adding an entropy variable, can only give a gradient flow, and in the Lotka--Volterra setting a genuine metriplectic structure needs at least two species together with the entropy variable.
\end{remark}

\begin{remark}\label{rem:leibniz}
Many dissipative systems can be written as $\dot y=(\Pi^\sharp-G^\sharp)(dV)$ for a single function $V$ on a manifold $P$, with $\Pi$ skew-symmetric and $G$ symmetric positive semidefinite. Such systems are sometimes called Leibniz or port-Hamiltonian systems; see, e.g., \cite{Guh07}. The function $V$ is then a Lyapunov function, but there is neither a conserved energy nor an entropy, so this is not a metriplectic system. If $\Pi$ is Poisson, such a system can always be closed to a metriplectic one on $Q=P\times\R$, with points $(y,s)$, by taking $E=V+s$, $S=s$, extending $\Pi$ with no components along $s$, and defining
\[
M(\alpha,\alpha')=G\big(\alpha_P-\alpha_s\,dV,\ \alpha'_P-\alpha'_s\,dV\big)
\]
for covectors $\alpha=\alpha_P+\alpha_s\,ds$ and $\alpha'=\alpha'_P+\alpha'_s\,ds$; the proof of Theorem~\ref{thm:metriplectic} applies verbatim. The construction is therefore general, and its content in a particular case lies in the existence of the decomposition with $\Pi$ Poisson and $G\ge0$. For LV systems this is where the log-canonical bivector and Volterra's condition enter.
\end{remark}

\section{Metriplectic structure of Lotka--Volterra systems}\label{sec:LV}

\subsection{Setting}

Consider the LV system on the positive orthant $\R^n_+$,
\begin{equation}\label{eq:LV}
\dot x_i=x_i\Big(r_i+\sum_{j=1}^nA_{ij}x_j\Big),\qquad i=1,\dots,n,
\end{equation}
and assume that it has an interior equilibrium $x^*\in\R^n_+$, that is, $r+Ax^*=0$. Then \eqref{eq:LV} reads
\begin{equation}\label{eq:LVeq}
\dot x_i=x_i\sum_jA_{ij}(x_j-x_j^*).
\end{equation}
Given a diagonal matrix $C=\diag(c_1,\dots,c_n)$ with $c_i>0$, let
\begin{equation}\label{eq:V}
V(x)=\sum_ic_i\big(x_i-x_i^*\log x_i\big),\qquad \partial_iV=c_i\,\frac{x_i-x_i^*}{x_i},
\end{equation}
be the associated Volterra function. We write $X=\diag(x_1,\dots,x_n)$,
\begin{equation}\label{eq:B}
B=AC^{-1},\qquad B^{\skw}=\tfrac12(B-B^\top),\qquad B^{\sym}=\tfrac12(B+B^\top),\qquad \Gamma=-B^{\sym}.
\end{equation}
Note that $C(B+B^\top)C=CA+A^\top C$, so that $\Gamma=-\tfrac12C^{-1}(CA+A^\top C)C^{-1}$.

\begin{lemma}\label{lem:decomposition}
System \eqref{eq:LVeq} can be written as
\begin{equation}\label{eq:decomposition}
\dot x=XBX\,\nabla V=XB^{\skw}X\,\nabla V+XB^{\sym}X\,\nabla V .
\end{equation}
\end{lemma}

\begin{proof}
By \eqref{eq:V}, $x_j-x_j^*=x_j\,\partial_jV/c_j$, and substituting in \eqref{eq:LVeq} gives $\dot x_i=\sum_jx_i(A_{ij}/c_j)x_j\,\partial_jV=(XBX\nabla V)_i$.
\end{proof}

The proof is purely algebraic and does not use the sign of $x^*$: identity \eqref{eq:decomposition} holds for any $x^*\in\R^n$ with $r+Ax^*=0$, with $V$ defined by \eqref{eq:V}. This will be used in Proposition~\ref{prop:Ito}.

\subsection{The Poisson bivector}

\begin{proposition}\label{prop:bivector}
Let $\beta$ be a constant skew-symmetric $n\times n$ matrix. Then
\begin{equation}\label{eq:Pi}
\Pi_\beta=\sum_{i<j}\beta_{ij}\,x_ix_j\,\partial_{x_i}\wedge\partial_{x_j},\qquad\text{i.e.}\quad \{x_i,x_j\}=\beta_{ij}x_ix_j,
\end{equation}
is a Poisson bivector on $\R^n_+$. In the logarithmic coordinates $u_i=\log x_i$ it is the constant bivector $\sum_{i<j}\beta_{ij}\,\partial_{u_i}\wedge\partial_{u_j}$. Consequently, its rank is $\operatorname{rank}\beta$, the functions $\sum_i\lambda_i\log x_i$ with $\lambda\in\ker\beta$ are Casimir functions, and $\Pi_\beta$ is unimodular with Liouville measure $\prod_idx_i/x_i=\prod_idu_i$.
\end{proposition}

\begin{proof}
Since $\partial_{x_i}=e^{-u_i}\partial_{u_i}$ and $x_ix_j=e^{u_i+u_j}$, we have $x_ix_j\,\partial_{x_i}\wedge\partial_{x_j}=\partial_{u_i}\wedge\partial_{u_j}$. A constant bivector satisfies the Jacobi identity, and the remaining assertions follow from linear algebra in the coordinates $u$, since Hamiltonian vector fields of a constant bivector are divergence-free with respect to $\prod_idu_i$.
\end{proof}

The reversible part of \eqref{eq:decomposition} is the Hamiltonian vector field of $V$ with respect to $\Pi_{B^{\skw}}$, since $(XB^{\skw}X)_{ij}=B^{\skw}_{ij}x_ix_j$. Since $C(B+B^\top)C=CA+A^\top C$, the classical conservative case of Volterra \cite{Vol31,Pla95}, in which $CA$ is skew-symmetric, corresponds to $B^{\sym}=0$.

\subsection{The metriplectic closure}

We add a variable $s\in\R$, which plays the role of the entropy of the environment, and work on $Q=\R^n_+\times\R$ with points $z=(x,s)$. For $i=1,\dots,n$ we consider the vectors
\[
w_i=\partial_iV\,e_s-e_i\in\R^{n+1},
\]
where $e_1,\dots,e_n,e_s$ is the standard basis associated with the coordinates $(x,s)$.

\begin{theorem}\label{thm:metriplectic}
Assume that \eqref{eq:LV} has an interior equilibrium $x^*$ and let $C$ be a positive diagonal matrix. Let
\begin{equation}\label{eq:ES}
E(x,s)=V(x)+s,\qquad S(x,s)=s,
\end{equation}
let $\Pi$ be the bivector $\Pi_{B^{\skw}}$ of Proposition~\ref{prop:bivector}, extended to $Q$ with no components along $s$, and let
\begin{equation}\label{eq:M}
M=\sum_{i,j=1}^nG_{ij}\,w_iw_j^{\top},\qquad G=X\Gamma X ,
\end{equation}
or, intrinsically, $M(\alpha,\alpha')=G(\alpha_x-\alpha_s\,dV,\,\alpha'_x-\alpha'_s\,dV)$ for covectors $\alpha=\alpha_x+\alpha_s\,ds$, $\alpha'=\alpha'_x+\alpha'_s\,ds$. Then $\Pi$ is a Poisson bivector, $\Pi^\sharp(dS)=0$ and $M^\sharp(dE)=0$. Moreover, $M$ is positive semidefinite if and only if
\begin{equation}\label{eq:Volterra}
CA+A^\top C\le0 .
\end{equation}
In that case $(\Pi,M,E,S)$ is a metriplectic system, the $x$-components of \eqref{eq:flow} are \eqref{eq:LV}, and
\[
\dot s=\nabla V^\top G\,\nabla V\ge0 .
\]
Finally, $\Pi\neq0$ if and only if $AC^{-1}$ is not symmetric, and $M\neq0$ if and only if $AC^{-1}$ is not skew-symmetric. We call the structure \emph{genuine} when both conditions hold.
\end{theorem}

\begin{proof}
The bivector is Poisson by Proposition~\ref{prop:bivector}, and it has no components along $s$, so $\Pi^\sharp(ds)=0$. Since $\nabla E=(\nabla V,1)$, we have $w_i^\top\nabla E=-\partial_iV+\partial_iV=0$ for every $i$, hence $M\nabla E=0$.

For $\zeta\in\R^{n+1}$ we have $\zeta^\top M\zeta=\eta^\top G\eta$ with $\eta_i=w_i^\top\zeta$, and the map $\zeta\mapsto\eta$ is onto (take $\zeta$ with vanishing $s$-component). This gives the intrinsic expression of $M$, and it shows that $M\ge0$ if and only if $G\ge0$, and that $M\neq0$ if and only if $G\neq0$. Since $X$ is invertible, $G\ge0$ if and only if $\Gamma\ge0$, i.e.\ $B^{\sym}\le0$, and this is equivalent to \eqref{eq:Volterra} because $C(B+B^\top)C=CA+A^\top C$.

Next, $\Pi^\sharp(dE)$ has $x$-components $XB^{\skw}X\nabla V$ and vanishing $s$-component, while
\[
M\nabla S=\sum_{i,j}G_{ij}\,(w_j^\top e_s)\,w_i=\sum_{i,j}G_{ij}\,\partial_jV\,w_i ,
\]
whose $x_i$-component is $-\sum_jG_{ij}\partial_jV=(XB^{\sym}X\nabla V)_i$ and whose $s$-component is $\nabla V^\top G\nabla V$. The identification with \eqref{eq:LV} follows from Lemma~\ref{lem:decomposition}. Finally, $\Pi\neq0$ if and only if $B^{\skw}\neq0$, that is, if and only if $B=AC^{-1}$ is not symmetric, and $M\neq0$ if and only if $G\neq0$, that is, if and only if $B^{\sym}\neq0$.
\end{proof}

Condition \eqref{eq:Volterra} is the classical condition ensuring that $V$ is a Lyapunov function for \eqref{eq:LV}, since $\dot V=\nabla V^\top XB^{\sym}X\nabla V=-\dot s$. Theorem~\ref{thm:metriplectic} gives a thermodynamic reading of this fact: the function $V$ dissipated by self-limitation is transferred to the environment variable $s$, and the total energy $E=V+s$ is conserved. Note that along the solutions $s=E_0-V(x)$, so the entropy variable is determined by the populations and the energy; the content of the construction lies in the decomposition, not in the extra variable (Remark~\ref{rem:leibniz}). The matrix $C$ is not unique, and different admissible choices give different bivectors and metrics.

The following proposition shows that condition \eqref{eq:Volterra} is necessary for every metriplectic structure with the energy and entropy \eqref{eq:ES}, whatever the bivector and the metric.

\begin{proposition}\label{prop:necessity}
Let $x^*\in\R^n_+$ be an interior equilibrium of \eqref{eq:LV}, let $C$ be a positive diagonal matrix, and let $E$, $S$ be given by \eqref{eq:ES}. Let $\Pi$ be any bivector field on $Q$ with $\Pi^\sharp(dS)=0$ and let $M$ be any symmetric positive semidefinite bilinear form on $T^*Q$ with $M^\sharp(dE)=0$. If the $x$-components of $\Pi^\sharp(dE)+M^\sharp(dS)$ coincide with the vector field \eqref{eq:LV}, then $CA+A^\top C\le0$.
\end{proposition}

\begin{proof}
Write the matrices of $\Pi$ and $M$ in blocks with respect to the coordinates $(x,s)$. Since $\Pi^\sharp(ds)=0$, the only nonzero block of $\Pi$ is the $xx$-block $\Pi_{xx}$, which is skew-symmetric, and the $x$-components of $\Pi^\sharp(dE)$ are $\Pi_{xx}\nabla V$. The condition $M\nabla E=0$, with $\nabla E=(\nabla V,1)$, gives $M_{xx}\nabla V+M_{xs}=0$, so the $x$-components of $M^\sharp(dS)=Me_s$ are $M_{xs}=-M_{xx}\nabla V$. By Lemma~\ref{lem:decomposition}, the hypothesis reads
\[
\Pi_{xx}\nabla V-M_{xx}\nabla V=XBX\nabla V .
\]
Contracting with $\nabla V$ and using the skew-symmetry of $\Pi_{xx}$ and $B^{\skw}$,
\[
-\nabla V^\top M_{xx}\nabla V=\nabla V^\top XB^{\sym}X\nabla V=y^\top B^{\sym}y,\qquad y=X\nabla V=C(x-x^*).
\]
The left-hand side is nonpositive, because $M_{xx}$ is a principal block of a positive semidefinite matrix. As $x$ ranges over $\R^n_+$, the vector $y$ ranges over $C(\R^n_+-x^*)$, which contains a neighbourhood of the origin because $x^*\in\R^n_+$. By homogeneity, $y^\top B^{\sym}y\le0$ for every $y\in\R^n$, i.e.\ $B^{\sym}\le0$, which is equivalent to \eqref{eq:Volterra}.
\end{proof}

Thus, the classical diagonal Lyapunov condition for Lotka–Volterra systems emerges naturally as a necessary condition for the existence of a metriplectic structure of the prescribed form. In this setting, the condition acquires a geometric interpretation through the positive semidefiniteness of the metric tensor.

The proof uses neither the Jacobi identity nor the precise form of $M$, only the positivity of its $xx$-block. On the other hand, we do not claim that $\Pi_{B^{\skw}}$ is the only bivector that can be used in a structure of this type, nor do we know whether an LV system violating \eqref{eq:Volterra} for every positive diagonal $C$ may admit a metriplectic structure with an energy not of Volterra type.

Condition \eqref{eq:Volterra} does not imply that the structure is genuine. If $A$ is diagonal, i.e.\ there is no interspecific interaction, then $AC^{-1}$ is diagonal for every $C$, so \eqref{eq:Volterra} holds whenever $A\le0$ but $\Pi=0$ for every admissible $C$. In this case the construction only yields $n$ uncoupled gradient flows, in agreement with Remark~\ref{rem:dimension}.

\subsection{Predator--prey with self-limitation}\label{sec:predator}

Consider
\begin{equation}\label{eq:predator}
\dot x=x(a-by-\varepsilon x),\qquad \dot y=y(-c+dx-\delta y),\qquad a,b,c,d,\varepsilon,\delta>0,
\end{equation}
whose interior equilibrium
\[
x^*=\frac{a\delta+bc}{bd+\varepsilon\delta},\qquad y^*=\frac{ad-c\varepsilon}{bd+\varepsilon\delta}
\]
lies in the positive quadrant if and only if $ad>c\varepsilon$. Here $A=\begin{psmallmatrix}-\varepsilon&-b\\ d&-\delta\end{psmallmatrix}$, and for $C=\diag(d,b)$ we get
\[
CA+A^\top C=\begin{pmatrix}-2d\varepsilon&0\\0&-2b\delta\end{pmatrix},
\]
so that the interaction between the two species is purely skew-symmetric. Up to a positive factor, this is the only choice of $C$ for which $CA+A^\top C$ is diagonal. We obtain
\[
V=d\,(x-x^*\log x)+b\,(y-y^*\log y),\qquad \{x,y\}=-xy,\qquad G=\diag\Big(\frac{\varepsilon x^2}{d},\ \frac{\delta y^2}{b}\Big),
\]
on $Q=\R^2_+\times\R$ with $z=(x,y,s)$, and
\[
\dot s=d\,\varepsilon\,(x-x^*)^2+b\,\delta\,(y-y^*)^2 .
\]
The Poisson part is the conservative LV system and the metric part is intraspecific competition. If $\varepsilon=\delta=0$ the system is Hamiltonian. Since $AC^{-1}=\begin{psmallmatrix}-\varepsilon/d&-1\\ 1&-\delta/b\end{psmallmatrix}$ is never symmetric, and is not skew-symmetric because $\varepsilon,\delta>0$, the structure is genuine for all values of the parameters.

For comparison, a single species following the logistic equation $\dot x=x(r-\varepsilon x)$, with $x^*=r/\varepsilon$, satisfies $\dot x=-\varepsilon x^2\,\Phi'(x)$ with $\Phi(x)=x-x^*\log x$. This is a gradient flow for the cometric $\varepsilon x^2$, i.e.\ for the Euclidean metric (scaled by $\varepsilon$) in the variable $u=\log x$, and by Remark~\ref{rem:dimension} it admits no Poisson part.

\section{Noise compatible with the structure}\label{sec:noise}

\subsection{Stochastic metriplectic equations}

Let $(\Pi,M,E,S)$ be a metriplectic system on $Q$, of dimension $N$, and write $M=\mathcal B\mathcal B^\top$, where $\mathcal B(z)$ is an $N\times m$ matrix, i.e.\ a linear map $\R^m\to T_zQ$. Let $W_t=(W^1_t,\dots,W^m_t)$ be a standard Brownian motion in $\R^m$ and $k>0$. The stochastic metriplectic (or stochastic GENERIC) equation \cite{Ott05,PS25} is the It\^o equation
\begin{equation}\label{eq:SDE}
dz=\Big(\Pi^\sharp(dE)+M^\sharp(dS)+k\,\dv_\mu M\Big)dt+\sqrt{2k}\,\mathcal B\,dW_t ,
\end{equation}
where $(\dv_\mu M)^a=\varrho^{-1}\partial_b(\varrho M^{ab})$ is the divergence with respect to a reference measure $\mu=\varrho\,dz$. Note that the Brownian motion lives in the auxiliary space $\R^m$ and not in $Q$; it enters the equation only through $\mathcal B$. The generator of \eqref{eq:SDE} is
\[
\mathcal L F=\langle dF,\Pi^\sharp(dE)+M^\sharp(dS)\rangle+k\,\varrho^{-1}\partial_a\big(\varrho M^{ab}\partial_bF\big),
\]
which is independent of the coordinates once $\mu$ is fixed, although the vector $\dv_\mu M$ is not; the role of the reference measure in a coordinate-free formulation is discussed in \cite{PS25}.

\begin{remark}[Choice of the reference measure]\label{rem:measure}
The reference measure is part of the data of \eqref{eq:SDE}. For $e^{S/k}\mu$, restricted to the level sets of $E$, to be invariant, the reversible part $\Pi^\sharp(dE)$ must preserve $\mu$, and we require that $\mu$ be a Liouville measure of $\Pi$, i.e.\ that it be preserved by all Hamiltonian vector fields of $\Pi$. Liouville measures are not unique: if $\mu_0$ is one, the Liouville measures with smooth positive density with respect to $\mu_0$ are exactly the measures $f\mu_0$ with $f>0$ a Casimir function, since the ratio of two such densities is annihilated by every Hamiltonian vector field. This freedom is absorbed by the entropy. Indeed, $\dv_{f\mu_0}M=\dv_{\mu_0}M+M^\sharp(d\log f)$, so equation \eqref{eq:SDE} with data $(S,f\mu_0)$ coincides with equation \eqref{eq:SDE} with data $(S+k\log f,\mu_0)$, and $(\Pi,M,E,S+k\log f)$ is again a metriplectic system, because $\log f$ is a Casimir function. Changing the Liouville measure therefore amounts to a redefinition of the entropy of order $k$ within the class of Casimir functions.

In the LV case the Casimir functions of $\Pi$ include $s$ and, when $B^{\skw}$ is singular, the functions $\lambda\cdot u$ with $\lambda\in\ker B^{\skw}$; the kernel is nontrivial, for instance, whenever $n$ is odd. These choices do affect the population equation: replacing $\mu_0$ by $f\mu_0$ adds to it the drift $k\,M^\sharp(d\log f)$, whose $u$-components are $k\,\Gamma\nabla_u\log f$ if $f=f(\lambda\cdot u)$ and $-k\,(\log f)'(s)\,\Gamma\nabla_uV$ if $f=f(s)$. We fix $\mu_0=\prod_idx_i/x_i\,ds$ (Proposition~\ref{prop:bivector}), which is the Lebesgue measure in the coordinates $(u,s)$, $u_i=\log x_i$, in which both $\Pi$ and $\Gamma$ are constant; equivalently, we fix the entropy to be exactly $s$. In the coordinates $(x,s)$ we have $M^{x_ix_j}=G_{ij}=\Gamma_{ij}x_ix_j$, and the $x_i$-component of $k\,\dv_{\mu_0}M$ is $k\,\Gamma_{ii}x_i$, which is precisely the It\^o correction of $dx_i=x_i\circ du_i$. The Euclidean measure $dx\,ds$ is not a Liouville measure of $\Pi$ unless $(1,\dots,1)\in\ker B^{\skw}$, and if it is used instead, a spurious drift $k\sum_jG_{ij}/x_j$ (which equals $k\,G_{ii}/x_i$ when $\Gamma$ is diagonal) appears in the equation for $x_i$, and $e^{-V/k}\mu_0$ is no longer stationary. One checks this directly in the example of Section~\ref{sec:predator}.
\end{remark}

\subsection{The Lotka--Volterra case}

From now on we use the coordinates $z=(u,s)$, $u_i=\log x_i$, on $Q$. In these coordinates $\Pi=\sum_{i<j}B^{\skw}_{ij}\,\partial_{u_i}\wedge\partial_{u_j}$ and, since $\partial_{u_i}V=x_i\,\partial_{x_i}V$, the metric of Theorem~\ref{thm:metriplectic} becomes
\[
M=\sum_{i,j}\Gamma_{ij}\,\omega_i\omega_j^\top,\qquad \omega_i=\partial_{u_i}V\,e_s-e_{u_i},
\]
where $\Gamma=-B^{\sym}\ge0$ is the \emph{constant} matrix of \eqref{eq:B}. We take $m=n$, one Brownian motion for each species, and $\mathcal B=(\omega_1,\dots,\omega_n)\,\Gamma^{1/2}$, an $(n+1)\times n$ matrix.

\begin{theorem}\label{thm:stochastic}
Assume \eqref{eq:Volterra} and let $k>0$. Consider equation \eqref{eq:SDE} for the metriplectic system of Theorem~\ref{thm:metriplectic}, written in the coordinates $(u,s)$, with $\mu=\mu_0=du\,ds$. Then:
\begin{enumerate}[label=\textup{(\alph*)}]
\item The noise-induced drift has no component along the populations: $(\dv_\mu M)^{u_i}=0$ and $(\dv_\mu M)^s=-\sum_i\Gamma_{ii}\,\partial^2_{u_i}V$.
\item The population components of \eqref{eq:SDE} are
\[
du=B\,\nabla_uV\,dt-\sqrt{2k}\,\Gamma^{1/2}dW_t ,
\]
which, in the original variables, is the Stratonovich equation
\begin{equation}\label{eq:Strat}
dx_i=x_i\Big(r_i+\sum_jA_{ij}x_j\Big)dt-x_i\sum_j\Sigma^{1/2}_{ij}\circ dW^j_t,\qquad \Sigma=2k\,\Gamma=-k\big(AC^{-1}+C^{-1}A^\top\big).
\end{equation}
\item The energy is conserved exactly: $dE=0$ almost surely.
\item The measure
\[
\pi(dx)\propto e^{-V(x)/k}\prod_i\frac{dx_i}{x_i}=\prod_ix_i^{\,c_ix_i^*/k-1}e^{-c_ix_i/k}\,dx_i ,
\]
which is the population marginal of the microcanonical measure $e^{S/k}\delta(E-E_0)\,\mu$, is stationary for \eqref{eq:Strat}. It is the product of the distributions $\mathrm{Gamma}\big(c_ix_i^*/k,\,c_i/k\big)$ (shape and rate).
\item If moreover $CA+A^\top C<0$, then \eqref{eq:Strat} has no explosion and $\pi$ is its unique stationary distribution on $\R^n_+$. In particular, in the stationary regime the populations are independent.
\end{enumerate}
\end{theorem}

\begin{proof}
(a) The entries $M^{u_iu_j}=\Gamma_{ij}$ are constant, and $M^{u_is}=-\sum_j\Gamma_{ij}\partial_{u_j}V$ and $M^{ss}=\sum_{i,j}\Gamma_{ij}\partial_{u_i}V\partial_{u_j}V$ do not depend on $s$. Hence $(\dv M)^{u_i}=\sum_j\partial_{u_j}M^{u_iu_j}+\partial_sM^{u_is}=0$, while $(\dv M)^s=\sum_i\partial_{u_i}M^{su_i}=-\sum_{i,j}\Gamma_{ij}\partial_{u_i}\partial_{u_j}V$, which reduces to the stated expression because $V$ is separable.

(b) The $u$-components of $\Pi^\sharp(dE)+M^\sharp(dS)$ are $B^{\skw}\nabla_uV+B^{\sym}\nabla_uV=B\nabla_uV$, as in the proof of Theorem~\ref{thm:metriplectic}, and the $u$-block of $\mathcal B$ is $-\Gamma^{1/2}$. Since the noise coefficient is constant in $u$, the It\^o and Stratonovich forms of this equation coincide, and $dx_i=x_i\circ du_i$ gives \eqref{eq:Strat}, because $(B\nabla_uV)_i=\sum_jA_{ij}(x_j-x_j^*)$. Replacing $W$ by $-W$ we may of course change the sign of the noise term.

(c) By It\^o's formula, $dE=\nabla E\cdot dz+k\,\tr(\nabla^2E\,M)\,dt$. The terms $\nabla E\cdot\Pi\nabla E$ and $\nabla E\cdot M\nabla S$ vanish as in Lemma~\ref{lem:laws}, and $\nabla E\cdot\mathcal B=0$ because $\omega_i\cdot\nabla E=0$. Finally,
\[
\nabla E\cdot\dv M=\sum_{a,b}\partial_aE\,\partial_bM^{ab}=\sum_b\partial_b\Big(\sum_aM^{ab}\partial_aE\Big)-\sum_{a,b}M^{ab}\partial_a\partial_bE=-\tr(\nabla^2E\,M),
\]
since $M\nabla E=0$, and this cancels the It\^o correction.

(d) The population components form a closed system $du=(J-\Gamma)\nabla_uV\,dt-\sqrt{2k}\,\Gamma^{1/2}dW_t$, with $J=B^{\skw}$ skew-symmetric and $\Gamma$ symmetric, both constant. The probability current of $e^{-V/k}du$ splits into the part generated by $\Gamma$, which vanishes by detailed balance, and the part $J\nabla_uV\,e^{-V/k}$, which is divergence-free because $J$ is constant and skew-symmetric. Hence $e^{-V/k}du$ is stationary. Returning to $x$, $du=\prod_idx_i/x_i$, and the separability of $V$ gives the product of Gamma densities. Integrating $e^{s/k}\delta(V+s-E_0)$ with respect to $s$ gives $e^{(E_0-V)/k}$, which is the statement about the microcanonical measure.

(e) If $CA+A^\top C<0$, then $\Gamma>0$, so the $u$-equation is a nondegenerate diffusion with additive noise and smooth drift. Let $\mathcal L_u$ be its generator. Since $J$ is skew-symmetric,
\[
\mathcal L_uV=-\nabla_uV^\top\Gamma\nabla_uV+k\sum_i\Gamma_{ii}\,c_i\,e^{u_i}
\le-\lambda\sum_ic_i^2\big(e^{u_i}-x_i^*\big)^2+k\sum_i\Gamma_{ii}\,c_i\,e^{u_i},
\]
where $\lambda>0$ is the smallest eigenvalue of $\Gamma$. The right-hand side is bounded above, and $V\to\infty$ as $|u|\to\infty$ because $x^*\in\R^n_+$. Hence $V$ is a Lyapunov function and the process does not explode \cite[Section~3.4]{Kha12}. A nondegenerate diffusion with smooth coefficients is strong Feller and irreducible, so it has at most one stationary distribution \cite[Sections~4.2--4.4]{Kha12}, which by (d) is $\pi$.
\end{proof}

The stationary measure depends on the interaction between species only through the equilibrium $x^*$: neither the reversible part nor the correlations of the noise create correlations between the populations, in the same way as a Hamiltonian flow does not alter a Gibbs distribution. The measure $\pi$ has the form of Kerner's canonical ensemble \cite{Ker57}, built on Volterra's integral of the conservative case and on the Liouville measure in logarithmic variables; here it arises instead as the stationary law of a dissipative stochastic dynamics, with the temperature $k$ fixed by the noise intensity. If $\Gamma$ is only positive semidefinite, which happens for instance when some species is not self-limited, the noise is degenerate and uniqueness would require a hypoellipticity argument that we do not pursue here.

\begin{remark}\label{rem:langevin}
The $u$-equation in the proof of (d) is a Langevin equation with a constant nonreversible perturbation $J\nabla_uV$, and the invariance of $e^{-V/k}du$ is the well-known fact that such perturbations preserve the Gibbs measure \cite{HHS93,RBS15}. From this point of view, the metriplectic structure identifies the coordinates $u=\log x$ in which the LV system takes this form, splits the interaction matrix into its reversible and irreversible parts, and fixes, through the fluctuation--dissipation relation, which noise has this property.
\end{remark}

\subsection{Consequences}

\begin{corollary}\label{cor:FDT}
The environmental noise compatible with the metriplectic structure has covariance
\[
\Sigma=-k\big(AC^{-1}+C^{-1}A^\top\big)=-k\,C^{-1}\big(CA+A^\top C\big)C^{-1}.
\]
Its diagonal entries are
\[
\Sigma_{ii}=\frac{2k\,|A_{ii}|}{c_i},\qquad i=1,\dots,n,
\]
so that the intensity of the noise acting on species $i$ is proportional to the ratio $|A_{ii}|/c_i$, with the same constant $k$ for all species. In particular, a species without self-limitation receives no noise. In general the noise is correlated between species, and the noises acting on different species are independent if and only if $CA+A^\top C$ is diagonal. In that case, if the interaction graph of $A$ (with an edge between $i\neq j$ whenever $A_{ij}\neq0$ or $A_{ji}\neq0$) is connected, $C$ is determined by $A$ up to a positive factor, so that the $n$ noise intensities are fixed by $A$ up to the single constant $k$.
\end{corollary}

\begin{proof}
The first formula follows from \eqref{eq:Strat}. The diagonal entries of $CA+A^\top C$ are $2c_iA_{ii}$, so $\Sigma_{ii}=-2kA_{ii}/c_i$, and \eqref{eq:Volterra} gives $A_{ii}\le0$. If $A_{ii}=0$, then $\Sigma_{ii}=|\Sigma^{1/2}e_i|^2=0$, so the $i$-th row of the symmetric matrix $\Sigma^{1/2}$ vanishes and no noise acts on species $i$. The matrix $\Sigma$ is diagonal if and only if $CA+A^\top C$ is, since $C$ is diagonal and invertible. In the diagonal case, $c_iA_{ij}+c_jA_{ji}=0$ for $i\neq j$; since $c_i,c_j>0$, $A_{ij}$ and $A_{ji}$ are either both zero or both nonzero, and in the latter case $c_j/c_i=-A_{ij}/A_{ji}$. If the interaction graph is connected, these relations determine all the ratios $c_j/c_1$.
\end{proof}

\begin{remark}\label{rem:family}
The fluctuation--dissipation relation selects a family of noises rather than a single one. The set of compatible covariances is
\[
\big\{-k\,C^{-1}(CA+A^\top C)C^{-1}\ :\ k>0,\ C \text{ positive diagonal with } CA+A^\top C\le0\big\},
\]
together with the choice of entropy discussed in Remark~\ref{rem:measure}, and each choice of $(C,k)$ gives a different stationary measure. When independent noises are required and the interaction graph is connected, the family reduces to a one-parameter family indexed by $k$, which makes the compatibility condition restrictive for $n\ge2$.
\end{remark}

\begin{proposition}\label{prop:Ito}
Consider the It\^o system
\[
dx_i=x_i\Big(\rho_i+\sum_jA_{ij}x_j\Big)dt+\sigma_i\,x_i\,dW^i_t,\qquad i=1,\dots,n,
\]
where $A$ is invertible, $A_{ii}<0$ for every $i$, $CA+A^\top C$ is diagonal for some positive diagonal matrix $C$, and $\sigma_i^2=2k|A_{ii}|/c_i$ for some $k>0$. Let $r_i=\rho_i-\sigma_i^2/2$ and $x^*_\sigma=-A^{-1}r$. Then:
\begin{enumerate}[label=\textup{(\roman*)}]
\item For every initial condition in $\R^n_+$ the solution exists for all $t\ge0$ and remains in $\R^n_+$.
\item If $x^*_\sigma\in\R^n_+$, the system has a unique stationary distribution on $\R^n_+$, which is the product of Gamma distributions of Theorem~\ref{thm:stochastic}\textup{(d)} computed with $x^*_\sigma$.
\item If some component of $x^*_\sigma$ is nonpositive, the system has no stationary distribution on $\R^n_+$.
\end{enumerate}
\end{proposition}

\begin{proof}
The It\^o equation $dx_i=\dots+\sigma_ix_i\,dW^i_t$ equals the Stratonovich one with drift reduced by $\tfrac12\sigma_i^2x_i$, so it coincides with \eqref{eq:Strat} for the rates $r_i$, possibly after replacing $W$ by $-W$. Under the assumptions, $CA+A^\top C$ is diagonal with entries $2c_iA_{ii}<0$, so it is negative definite, $\Gamma$ is diagonal with entries $|A_{ii}|/c_i>0$, and there is $\lambda>0$ such that $x^\top CAx=\tfrac12x^\top(CA+A^\top C)x\le-\lambda|x|^2$.

(i) In the variables $u=\log x$ the equation reads $du_i=(r_i+(Ax)_i)\,dt+\sigma_i\,dW^i_t$. Let $\Psi(u)=\sum_ic_i(e^{u_i}-u_i)$, which tends to $\infty$ as $|u|\to\infty$. Then
\[
\mathcal L_u\Psi=\sum_ic_i(x_i-1)\big(r_i+(Ax)_i\big)+\tfrac12\sum_i\sigma_i^2c_ix_i
=x^\top CAx+\ell(x)\le-\lambda|x|^2+\ell(x),
\]
where $\ell$ is an affine function of $x$. Hence $\mathcal L_u\Psi$ is bounded above on $\R^n$, and the process does not explode \cite[Section~3.4]{Kha12}.

(ii) If $x^*_\sigma\in\R^n_+$, then $r+Ax^*_\sigma=0$ and Theorem~\ref{thm:stochastic}\textup{(d)--(e)} applies with $x^*=x^*_\sigma$.

(iii) Let $V_\sigma$ be defined by \eqref{eq:V} with $x^*_\sigma$ in place of $x^*$. By the remark after Lemma~\ref{lem:decomposition}, the $u$-equation is still $du=(J-\Gamma)\nabla_uV_\sigma\,dt-\sqrt{2k}\,\Gamma^{1/2}dW_t$, and the computation in the proof of Theorem~\ref{thm:stochastic}\textup{(d)} shows that $\varrho_\sigma=e^{-V_\sigma/k}$ is a smooth positive solution of the stationary Fokker--Planck equation. Its integral over $\R^n$ is the product of the integrals $\int_0^\infty x_i^{c_ix^*_{\sigma,i}/k-1}e^{-c_ix_i/k}\,dx_i$, and the one corresponding to a nonpositive component of $x^*_\sigma$ diverges at the origin, so $\varrho_\sigma$ is not integrable. Suppose that the system has a stationary distribution on $\R^n_+$. The $u$-process is a nondegenerate diffusion with smooth coefficients and, by (i), it does not explode, so the existence of a stationary distribution implies that it is positive recurrent \cite[Sections~4.3--4.4]{Kha12}. For a recurrent diffusion, the positive solutions of the stationary Fokker--Planck equation are unique up to a constant factor \cite[Theorems~4.3.3, 4.3.4 and 4.8.6]{Pin95}, so $\varrho_\sigma$ would be a multiple of the density of the stationary distribution, which contradicts the non-integrability of $\varrho_\sigma$.
\end{proof}

\begin{example}[The logistic equation]
For $n=1$ and $dx=x(\rho-\varepsilon x)\,dt+\sigma x\,dW_t$ (It\^o), we have $x^*_\sigma=(\rho-\sigma^2/2)/\varepsilon$ and $k=\sigma^2c/(2\varepsilon)$, so that
\[
\pi(x)\propto x^{\frac{2\rho}{\sigma^2}-2}\,e^{-2\varepsilon x/\sigma^2},
\]
which is the well-known Gamma distribution. By Proposition~\ref{prop:Ito}, a stationary distribution on $(0,\infty)$ exists if and only if $x^*_\sigma>0$, which is the classical persistence threshold $\rho>\sigma^2/2$ \cite{MSR03}. There is no Poisson part in this case (Remark~\ref{rem:dimension}), and for $n=1$ the compatibility condition is empty, since every $\sigma$ determines a value of $k$. The example nevertheless shows the role of each ingredient: the metric $\varepsilon x^2$, the temperature $k$, and the persistence threshold as the normalizability of the Gibbs measure.
\end{example}

\begin{example}[Predator--prey]
For system \eqref{eq:predator} with It\^o noise terms $\sigma_1x\,dW^1_t$ and $\sigma_2y\,dW^2_t$, where $W^1,W^2$ are independent standard Brownian motions, compatibility requires
\[
\frac{\sigma_1^2\,d}{\varepsilon}=\frac{\sigma_2^2\,b}{\delta}=2k .
\]
Let $(x^*_\sigma,y^*_\sigma)$ be the equilibrium of \eqref{eq:predator} with $a$ replaced by $a-\sigma_1^2/2$ and $c$ by $c+\sigma_2^2/2$. Then $y^*_\sigma>0$ if and only if
\begin{equation}\label{eq:coexistence}
\Big(a-\frac{\sigma_1^2}{2}\Big)d>\Big(c+\frac{\sigma_2^2}{2}\Big)\varepsilon ,
\end{equation}
and this condition implies $a>\sigma_1^2/2$, hence $x^*_\sigma>0$. By Proposition~\ref{prop:Ito}, under the compatibility condition the system has a stationary distribution in the open quadrant if and only if \eqref{eq:coexistence} holds. In that case the stationary distribution is unique, and in the stationary regime $x$ and $y$ are independent, with
\[
x\sim\mathrm{Gamma}\Big(\frac{2\varepsilon x^*_\sigma}{\sigma_1^2},\frac{2\varepsilon}{\sigma_1^2}\Big),\qquad y\sim\mathrm{Gamma}\Big(\frac{2\delta y^*_\sigma}{\sigma_2^2},\frac{2\delta}{\sigma_2^2}\Big).
\]
\end{example}

\begin{remark}\label{rem:invasion}
The threshold \eqref{eq:coexistence} does not depend on the compatibility condition. For arbitrary $\sigma_1,\sigma_2>0$, the persistence of the predator is governed by its invasion rate at the stationary distribution of the prey alone \cite{HN18}. The prey alone follows the stochastic logistic equation, whose stationary distribution, for $a>\sigma_1^2/2$, has mean $(a-\sigma_1^2/2)/\varepsilon$, so the invasion rate of the predator is
\[
-c-\frac{\sigma_2^2}{2}+d\,\frac{a-\sigma_1^2/2}{\varepsilon},
\]
which is positive exactly when \eqref{eq:coexistence} holds. What compatibility provides is not the threshold but the explicit form of the stationary law and the independence of the two populations in the stationary regime.
\end{remark}

\section{Interactions are invisible at stationarity}\label{sec:blind}

By Theorem~\ref{thm:stochastic}, the stationary law of a compatible system with $CA+A^\top C<0$ is
\[
\bigotimes_{i=1}^n\mathrm{Gamma}\big(c_ix_i^*/k,\,c_i/k\big),
\]
which depends on the interaction matrix only through the equilibrium $x^*$ and on the structure only through the matrix $C/k$. The next theorem makes precise how much freedom this leaves.

\begin{theorem}\label{thm:blind}
Let $\alpha,\beta\in\R^n_+$, let $\pi=\bigotimes_i\mathrm{Gamma}(\alpha_i,\beta_i)$ (shape and rate), and write $m_i=\alpha_i/\beta_i$ and $D=\diag(\beta_1,\dots,\beta_n)$. Let
\[
\mathcal A_D=\big\{A\in\R^{n\times n}\ :\ DA+A^\top D<0\big\}.
\]
\begin{enumerate}[label=\textup{(\alph*)}]
\item For every $A\in\mathcal A_D$, let $r=-Am$ and $\Sigma_A=-(AD^{-1}+D^{-1}A^\top)$. Then, for every $k>0$, the Stratonovich system \eqref{eq:Strat} with data $(r,A,\Sigma_A)$ is the compatible stochastic system of Theorem~\ref{thm:stochastic} with $C=kD$ and temperature $k$, and $\pi$ is its unique stationary distribution.
\item Conversely, if the compatible system of Theorem~\ref{thm:stochastic}, with $CA+A^\top C<0$, has stationary distribution $\pi$, then $C/k=D$, $x^*=m$ and $A\in\mathcal A_D$.
\item A matrix $A$ belongs to $\mathcal A_D$ if and only if
\[
A=\big(\Omega-\tfrac12\Sigma\big)D,\qquad \Omega^\top=-\Omega,\qquad \Sigma=\Sigma^\top>0,
\]
and in that case $\Sigma_A=\Sigma$ and $\Omega$ is the reversible part $kB^{\skw}$ of Theorem~\ref{thm:metriplectic}. In particular, $\mathcal A_D$ is an open cone of dimension $n^2$, and for fixed $\pi$ and fixed noise covariance $\Sigma$ the stationary law is the same for every skew-symmetric matrix $\Omega$.
\end{enumerate}
\end{theorem}

\begin{proof}
(a) With $C=kD$ we have $CA+A^\top C=k(DA+A^\top D)<0$, and $-k(AC^{-1}+C^{-1}A^\top)=-(AD^{-1}+D^{-1}A^\top)=\Sigma_A$, which is the covariance of \eqref{eq:Strat}. The equilibrium is $x^*=m\in\R^n_+$, and the Gamma parameters of Theorem~\ref{thm:stochastic}\textup{(d)} are $c_ix_i^*/k=\beta_im_i=\alpha_i$ and $c_i/k=\beta_i$. Uniqueness follows from Theorem~\ref{thm:stochastic}\textup{(e)}.

(b) A product of Gamma distributions determines its shape and rate parameters, so $c_i/k=\beta_i$ and $c_ix_i^*/k=\alpha_i$, i.e.\ $C/k=D$ and $x^*=m$, and $CA+A^\top C<0$ gives $A\in\mathcal A_D$.

(c) Write $B_D=AD^{-1}$. Since $DA+A^\top D=D(B_D+B_D^\top)D$, we have $A\in\mathcal A_D$ if and only if the symmetric part of $B_D$ is negative definite. Writing $B_D=\Omega-\frac12\Sigma$ with $\Omega$ its skew-symmetric part and $-\frac12\Sigma$ its symmetric part gives the parametrization, and $\Sigma_A=-(B_D+B_D^\top)=\Sigma$. Since $C=kD$, $B_D=kAC^{-1}=kB$, so $\Omega=kB^{\skw}$. The parameters $(\Omega,\Sigma)$ range over an open subset of a vector space of dimension $\frac12n(n-1)+\frac12n(n+1)=n^2$, and the set is invariant under multiplication by positive scalars.
\end{proof}

The theorem says that the stationary law sees the Volterra function, through $m=x^*$ and $D=C/k$, and the noise sees the metric part, through $\Sigma=2k\Gamma$, but neither sees the Poisson part. Since the noise covariance $\Sigma$ is not determined by $\pi$ either, the stationary law alone does not even determine the symmetric part of the interaction.

\begin{corollary}\label{cor:twin}
Let $A\in\mathcal A_D$ and suppose that $\Sigma_A=\diag(\sigma_1^2,\dots,\sigma_n^2)$ is diagonal. Then the interacting system with data $(r,A,\Sigma_A)$ and the system of $n$ uncoupled stochastic logistic equations
\[
dx_i=x_i\big(\tfrac12\sigma_i^2\beta_i\,(m_i-x_i)\big)dt+\sigma_i\,x_i\circ dW^i_t,\qquad i=1,\dots,n,
\]
are driven by the same noise and have the same unique stationary distribution $\pi$.
\end{corollary}

\begin{proof}
The uncoupled system corresponds to $\Omega=0$ in Theorem~\ref{thm:blind}\textup{(c)}, i.e.\ $A'=-\frac12\Sigma_AD=-\frac12\diag(\sigma_i^2\beta_i)$, with $r'=-A'm$ and $\Sigma_{A'}=\Sigma_A$.
\end{proof}

\begin{example}[Predator--prey]\label{ex:pp-twin}
For system \eqref{eq:predator} with compatible noise (Section~\ref{sec:predator}), $D=\diag(d,b)/k$, $\Sigma=2k\diag(\varepsilon/d,\delta/b)$ and $\Omega=\begin{psmallmatrix}0&-k\\ k&0\end{psmallmatrix}$. Its uncoupled twin is the pair of logistic equations $\dot x=\varepsilon x(x^*-x)$, $\dot y=\delta y(y^*-y)$, driven by the same noise. The predator--prey interaction, however strong, leaves no trace in the stationary distribution of the populations.
\end{example}

\begin{remark}\label{rem:SLM}
In \cite{Gri20}, the agreement of abundance fluctuations in microbial communities with the stationary Gamma law of the stochastic logistic model, in which species do not interact, supports a description in which environmental fluctuations are the main determinant of abundance variation. Theorem~\ref{thm:blind} and Corollary~\ref{cor:twin} show that, within the class of LV communities with compatible noise, the same stationary Gamma law can arise from different interaction structures, including reversible and symmetric interactions consistent with the noise. Thus, observing this stationary law does not rule out interspecific interactions, nor does it allow them to be identified within this class. Distinguishing between these possibilities requires dynamical information, to which we now turn.
\end{remark}

\begin{remark}[Relation with Gibbs measures for symmetric interactions]\label{rem:disordered}
In the statistical physics of disordered ecosystems \cite{Bun17,BBC18,ARCB21,GH25}, the stochastic LV system with demographic noise $\sqrt{2Tx_i}\,dW^i_t$ and a symmetric interaction matrix has a stationary Boltzmann distribution, because the drift is then a gradient with respect to the metric selected by the noise. In our language this is a case without Poisson part. Theorems~\ref{thm:stochastic} and~\ref{thm:blind} provide a different mechanism, based on environmental noise: the interaction may have an arbitrary skew-symmetric part with respect to the weights $C$, and the stationary law is still an explicit Gibbs measure, provided that the noise satisfies the fluctuation--dissipation relation. The two settings select different reference measures: the log-canonical bivector requires $\prod_idx_i/x_i$ (Remark~\ref{rem:measure}), which is not the measure associated with demographic noise. Whether a class of nonsymmetric interactions with an explicit stationary law exists for demographic noise is, as far as we know, open.
\end{remark}

\section{Interactions are visible in time}\label{sec:time}

\subsection{Moment identities}

We now consider a general Stratonovich LV system with constant noise covariance,
\begin{equation}\label{eq:general}
dx_i=x_i\Big(r_i+\sum_jA_{ij}x_j\Big)dt+x_i\sum_j\Lambda_{ij}\circ dW^j_t,\qquad Q=\Lambda\Lambda^\top,
\end{equation}
which in the variables $u=\log x$ reads $du=(r+Ax)\,dt+\Lambda\,dW_t$. We do not assume compatibility. For a stationary solution we write $\bar x=\mathbb E[x]$, $\bar u=\mathbb E[u]$,
\[
K=\Cov(x,u),\quad K_{ij}=\Cov(x_i,u_j),\qquad R(\tau)=\mathbb E\big[(u(t+\tau)-\bar u)(u(t)-\bar u)^\top\big],\quad \tau\ge0 .
\]

\begin{theorem}\label{thm:moments}
Let $x(t)$ be a stationary solution of \eqref{eq:general} with $\mathbb E|x|^2<\infty$ and $\mathbb E|u|^2<\infty$. Then:
\begin{enumerate}[label=\textup{(\alph*)}]
\item $r+A\bar x=0$; in particular, if $A$ is invertible, $\bar x=x^*$.
\item The function $R$ is differentiable from the right at $\tau=0$, and $\dot R(0^+)=A\,K$.
\item $\dot R(0^+)+\dot R(0^+)^\top=-Q$.
\end{enumerate}
\end{theorem}

\begin{proof}
Let $b(u)=r+Ax$ and $\Delta=u(\tau)-u(0)=\int_0^\tau b(u(s))\,ds+\Lambda W_\tau$, with $W$ the Brownian increment after time $0$, which is independent of $u(0)$.

(a) By stationarity $\mathbb E[\Delta]=0$, and $\mathbb E[\Lambda W_\tau]=0$, so $\tau\,\mathbb E[b]=0$.

(b) Multiplying $\Delta$ by $(u(0)-\bar u)^\top$ and taking expectations, the noise term vanishes by independence, and by (a)
\[
R(\tau)-R(0)=\int_0^\tau A\,\mathbb E\big[(x(s)-\bar x)(u(0)-\bar u)^\top\big]\,ds .
\]
The integrand is continuous in $s$: the paths are continuous, and the family $\{|x(s)|\,|u(0)|\}_{s\ge0}$ is uniformly integrable, because it is dominated by $\frac12(|x(s)|^2+|u(0)|^2)$ and the variables $|x(s)|^2$ are identically distributed. Hence $\dot R(0^+)$ exists and equals the integrand at $s=0$, which is $AK$.

(c) By stationarity, $\mathbb E[(u(\tau)-\bar u)(u(\tau)-\bar u)^\top]=R(0)$. Expanding $u(\tau)-\bar u=(u(0)-\bar u)+\Delta$ gives $R(\tau)+R(\tau)^\top-2R(0)=-\mathbb E[\Delta\Delta^\top]$. Since $\mathbb E|\int_0^\tau b\,ds|^2\le\tau^2\,\mathbb E|b|^2$ and $\mathbb E[\Lambda W_\tau(\Lambda W_\tau)^\top]=Q\tau$, the Cauchy--Schwarz inequality gives $\mathbb E[\Delta\Delta^\top]=Q\tau+O(\tau^{3/2})$. Dividing by $\tau$ and using (b) gives the claim.
\end{proof}

\begin{remark}\label{rem:moments-ito}
For It\^o noise $x_i\sum_j\Lambda_{ij}\,dW^j$, the drift of $u$ is $\rho-\frac12\diag(Q)+Ax$, and Theorem~\ref{thm:moments} holds with $r=\rho-\frac12\diag(Q)$. Part (a) is the stochastic counterpart of the classical fact that time averages of persistent deterministic LV trajectories equal the equilibrium \cite{Vol31,GMM71}.
\end{remark}

\begin{remark}\label{rem:MAR}
In first-order multivariate autoregressive (MAR(1)) models \cite{Ives03}, the log-abundances are modelled by a linear stochastic recursion around equilibrium, and the interaction matrix is estimated by regression of increments on current values. Part (b) of Theorem~\ref{thm:moments} is an exact, nonlinear counterpart for LV systems in continuous time: the regression of the increments of $\log x$ on $\log x$ involves the LV matrix $A$ itself, not its linearization $AX^*$, through the covariance $\Cov(x,\log x)$, and no small-noise approximation is needed.
\end{remark}

\begin{corollary}\label{cor:identify}
Under the assumptions of Theorem~\ref{thm:moments}, if $K$ is invertible, then
\[
A=\dot R(0^+)\,K^{-1},\qquad Q=-\big(\dot R(0^+)+\dot R(0^+)^\top\big),\qquad r=-A\bar x ,
\]
so the system is determined by the one-time moments $\bar x$, $K$ and by $\dot R(0^+)$. For the compatible system of Theorem~\ref{thm:blind}, $K=D^{-1}=\diag\big(\Var(x_i)/\mathbb E[x_i]\big)$, and
\[
\dot R(0^+)=AD^{-1}=kAC^{-1}=\Omega-\tfrac12\Sigma,\qquad\text{i.e.}\qquad A_{ij}=\dot R_{ij}(0^+)\,\frac{\mathbb E[x_j]}{\Var(x_j)} .
\]
In particular, the reversible part $\Omega$ of the interaction is the antisymmetric part of $\dot R(0^+)$, and the lagged covariances satisfy
\[
R(\tau)-R(\tau)^\top=2\tau\,\Omega+o(\tau),\qquad \tau\to0^+ .
\]
\end{corollary}

\begin{proof}
The first part follows from Theorem~\ref{thm:moments}. In the compatible case the populations are independent with $x_i\sim\mathrm{Gamma}(\alpha_i,\beta_i)$, so $K_{ij}=0$ for $i\ne j$, while $\Cov(x_i,\log x_i)=1/\beta_i$, since $\mathbb E[x_i\log x_i]=(\alpha_i/\beta_i)\,\mathbb E[\log x_i]+1/\beta_i$ for the Gamma distribution, and $\Var(x_i)/\mathbb E[x_i]=1/\beta_i$. Alternatively, $K=kC^{-1}$ follows from the integration by parts $\mathbb E[\partial_{u_i}V\,u_j]=k\,\delta_{ij}$ with respect to $e^{-V/k}du$, since $\partial_{u_i}V=c_i(x_i-x_i^*)$. The remaining formulas follow from Theorem~\ref{thm:blind}\textup{(c)}.
\end{proof}

Theorem~\ref{thm:moments} is an elementary consequence of It\^o's formula, and it holds for any noise covariance. Its interest here is that, combined with Theorem~\ref{thm:blind}, it shows exactly where each part of the interaction becomes visible: the symmetric part of $\dot R(0^+)$ is fixed by the noise, and the information that the stationary law misses is contained in its antisymmetric part, i.e.\ in the time asymmetry of the cross-covariances of the log-abundances.

\subsection{Entropy production}

The time asymmetry can be quantified by the stationary entropy production rate. For a stationary diffusion $du=b\,dt+\Lambda\,dW$ with $Q$ invertible, stationary density $\pi$ and stationary probability current $j=b\pi-\frac12Q\nabla\pi$, it is
\[
e_p=\int\frac{j^\top(\tfrac12Q)^{-1}j}{\pi}\,du ,
\]
which equals the relative entropy per unit time between the stationary path measure and its time reversal, and vanishes if and only if the process is reversible \cite{JQQ04}.

\begin{theorem}\label{thm:ep}
For the compatible system of Theorem~\ref{thm:blind}, with $A=(\Omega-\frac12\Sigma)D$,
\[
e_p=2\sum_{i=1}^n\alpha_i\,\big(\Omega^\top\Sigma^{-1}\Omega\big)_{ii}
=\tr\!\Big(J^\top\Gamma^{-1}J\,\diag(c_1x_1^*,\dots,c_nx_n^*)\Big),\qquad J=B^{\skw}.
\]
In particular $e_p$ does not depend on $k$ once $A$ and $C$ are fixed, and $e_p=0$ if and only if $\Omega=0$, i.e.\ if and only if the Poisson part of the metriplectic structure vanishes.
\end{theorem}

\begin{proof}
In the variables $u$, $b=A(x-m)=(\Omega-\frac12\Sigma)D(x-m)$, and the stationary density $\pi\propto e^{-V/k}$ satisfies $\nabla_u\log\pi=-D(x-m)$. Hence $b=-\Omega\nabla_u\log\pi+\frac12\Sigma\nabla_u\log\pi$, and $j=b\pi-\frac12\Sigma\nabla\pi=-\Omega\nabla_u\pi$. Therefore
\[
e_p=2\,\mathbb E\big[\nabla_u\log\pi^\top\,\Omega^\top\Sigma^{-1}\Omega\,\nabla_u\log\pi\big]=2\tr\big(\Omega^\top\Sigma^{-1}\Omega\,I_\pi\big),
\]
where $I_\pi=\mathbb E[\nabla_u\log\pi\,\nabla_u\log\pi^\top]=D\Cov(x)D=\diag(\beta_i^2\,\alpha_i/\beta_i^2)=\diag(\alpha_i)$. With $\Omega=kJ$, $\Sigma=2k\Gamma$ and $\alpha_i=c_ix_i^*/k$ one obtains the second expression, which does not contain $k$. Since $\Sigma>0$ and $\alpha_i>0$, $e_p=0$ if and only if $\Omega=0$.
\end{proof}

\begin{example}
For the predator--prey model of Example~\ref{ex:pp-twin},
\[
e_p=bd\Big(\frac{x^*}{\delta}+\frac{y^*}{\varepsilon}\Big),
\]
which grows without bound as the self-limitation vanishes, in agreement with the fact that the limit is a Hamiltonian system with degenerate noise. For the uncoupled twin of Corollary~\ref{cor:twin}, $e_p=0$.
\end{example}

\begin{remark}\label{rem:duality}
The deterministic entropy production of the metriplectic system, $\dot s=M(dS,dS)$, comes from the metric part. In the stationary regime of the stochastic system, the metric part is in detailed balance with the noise, and the whole irreversibility of the population process, as measured by $e_p$, is carried by the Poisson part. The two parts of the structure thus play opposite roles in the deterministic relaxation and in the stationary fluctuations.
\end{remark}

\section{Incompatible noise and correlations}\label{sec:incompatible}

We return to system \eqref{eq:general} with an arbitrary positive definite covariance $Q=\Lambda\Lambda^\top$ and an interior equilibrium $x^*$, and assume that $CA+A^\top C<0$ for some positive diagonal $C$, so that $x^*$ is globally stable for the deterministic system. We do not assume compatibility. For a stationary solution with $\mathbb E|x|^2+\mathbb E|u|^2<\infty$ we write $K=\Cov(x,u)$ as in Section~\ref{sec:time}, and we split it as $K=K_d+K_o$ into its diagonal part $K_d$ and its off-diagonal part $K_o$. Note that $K$ is not symmetric in general: $K_{ij}=\Cov(x_i,\log x_j)$ and $K_{ji}=\Cov(x_j,\log x_i)$ are different quantities. Existence of stationary solutions for incompatible noise is guaranteed, for instance, by the invasion criteria of \cite{HN18}.

\subsection{An exact identity}

\begin{lemma}\label{lem:Kii}
For any stationary solution as above, $K_{ii}=\Cov(x_i,\log x_i)>0$ for every $i$.
\end{lemma}

\begin{proof}
Let $x_i'$ be an independent copy of $x_i$. Since $\log$ is increasing, $(x_i-x_i')(\log x_i-\log x_i')\ge0$, with equality only if $x_i=x_i'$, and $2K_{ii}=\mathbb E[(x_i-x_i')(\log x_i-\log x_i')]$. Moreover, $x_i$ is not almost surely constant: otherwise, by stationarity, $u_i(t)=\log x_i(t)$ would be almost surely constant in $t$, whereas its quadratic variation is $\langle u_i\rangle_t=Q_{ii}\,t$ with $Q_{ii}>0$. Hence $K_{ii}>0$.
\end{proof}

\begin{theorem}\label{thm:exact}
For any stationary solution of \eqref{eq:general} with $\mathbb E|x|^2+\mathbb E|u|^2<\infty$,
\begin{equation}\label{eq:exact-lyap}
AK+K^\top A^\top+Q=0 .
\end{equation}
Moreover, $K_o=0$ if and only if $Q$ is compatible, that is,
\[
Q=-k\big(AC'^{-1}+C'^{-1}A^\top\big)
\]
for some $k>0$ and some positive diagonal matrix $C'$. In that case $C'A+A^\top C'<0$, the stationary solution is unique, and $K=K_d=k\,C'^{-1}$.
\end{theorem}

\begin{proof}
Identity \eqref{eq:exact-lyap} is Theorem~\ref{thm:moments}\textup{(b)--(c)}: $\dot R(0^+)=AK$ and $\dot R(0^+)+\dot R(0^+)^\top=-Q$.

Suppose that $K_o=0$. By Lemma~\ref{lem:Kii}, $\Delta=K_d$ is positive diagonal, and \eqref{eq:exact-lyap} reads $Q=-(A\Delta+\Delta A^\top)$. Given $k>0$, let $C'=k\Delta^{-1}$. Then $-k(AC'^{-1}+C'^{-1}A^\top)=-(A\Delta+\Delta A^\top)=Q$, and
\[
C'A+A^\top C'=k\,\Delta^{-1}\big(A\Delta+\Delta A^\top\big)\Delta^{-1}=-k\,\Delta^{-1}Q\,\Delta^{-1}<0 .
\]
Conversely, suppose that $Q$ is compatible for $(k,C')$. Since $Q>0$, the same computation gives $C'A+A^\top C'=-k^{-1}C'QC'<0$, so Theorem~\ref{thm:stochastic}\textup{(e)} applies with the matrix $C'$: the stationary solution is unique and its law is a product of Gamma distributions. The populations are then independent, so $K_o=0$, and $K=kC'^{-1}$ by Corollary~\ref{cor:identify}.
\end{proof}

Identity \eqref{eq:exact-lyap} is an exact, nonlinear analogue of the Lyapunov equation of the linear-noise approximation (Section~\ref{sec:linear}), with $\Cov(x,\log x)$ in place of $X^*\Cov(u)$; it holds for every noise intensity. It does not determine $K$, since only the symmetric part of $AK$ appears, but it constrains it enough to give the following consequences.

\begin{corollary}\label{cor:independence}
Under the assumptions of Theorem~\ref{thm:exact}, the following are equivalent:
\begin{enumerate}[label=\textup{(\roman*)}]
\item the stationary populations $x_1,\dots,x_n$ are independent;
\item $\Cov(x_i,\log x_j)=0$ for all $i\neq j$;
\item the noise covariance $Q$ is compatible with a metriplectic structure of the form of Theorem~\ref{thm:metriplectic}.
\end{enumerate}
In that case the stationary law is the product of Gamma distributions of Theorem~\ref{thm:stochastic}, with rates $1/K_{ii}$.
\end{corollary}

\begin{proof}
(i) implies (ii) trivially, (ii) implies (iii) and (iii) implies (i) by Theorem~\ref{thm:exact} and Theorem~\ref{thm:stochastic}. The rates are $c_i'/k=1/K_{ii}$.
\end{proof}

Corollary~\ref{cor:independence} is a converse of Theorem~\ref{thm:stochastic}: compatibility is not only sufficient but also necessary for the stationary populations of an LV system with environmental noise to be independent. Combined with Theorem~\ref{thm:blind}, it shows that, among LV systems with constant noise covariance, the product form of the stationary law characterizes exactly the stochastic metriplectic ones.

\subsection{A lower bound for the correlations}

The distance of $Q$ to the compatible family controls the size of the correlations quantitatively. Let
\[
\mathcal Q_A=\big\{-(A\Delta+\Delta A^\top)\ :\ \Delta\ \text{diagonal}\big\},
\]
a linear subspace of the space of symmetric matrices, of dimension at most $n$, which contains every compatible covariance. We use the Frobenius norm $\|\cdot\|_F$ and the operator norm $\|\cdot\|_{\mathrm{op}}$, and write $d_F(Q,\mathcal Q_A)$ for the Frobenius distance of $Q$ to $\mathcal Q_A$.

\begin{proposition}\label{prop:bound}
Under the assumptions of Theorem~\ref{thm:exact},
\[
\|K_o\|_F\ \ge\ \frac{d_F(Q,\mathcal Q_A)}{2\,\|A\|_{\mathrm{op}}} .
\]
\end{proposition}

\begin{proof}
By \eqref{eq:exact-lyap}, $Q+AK_d+K_dA^\top=-(AK_o+K_o^\top A^\top)$. The left-hand side is $Q-q$ with $q=-(AK_d+K_dA^\top)\in\mathcal Q_A$, so its Frobenius norm is at least $d_F(Q,\mathcal Q_A)$. The right-hand side has Frobenius norm at most $2\|AK_o\|_F\le2\|A\|_{\mathrm{op}}\|K_o\|_F$.
\end{proof}

The bound holds for every noise intensity, without linearization, and it is explicit: $d_F(Q,\mathcal Q_A)$ is the residual of a linear least-squares problem in the $n$ diagonal entries of $\Delta$.

\begin{example}[Predator--prey]\label{ex:pp-bound}
For system \eqref{eq:predator} and $\Delta=\diag(\delta_1,\delta_2)$,
\[
-(A\Delta+\Delta A^\top)=\begin{pmatrix}2\varepsilon\delta_1 & b\delta_2-d\delta_1\\ b\delta_2-d\delta_1 & 2\delta\,\delta_2\end{pmatrix}.
\]
Identifying a symmetric $2\times2$ matrix $\begin{psmallmatrix}p&q\\ q&s\end{psmallmatrix}$ with $(p,s,\sqrt2\,q)\in\R^3$, which is an isometry for the Frobenius norm, $\mathcal Q_A$ is the plane spanned by $(2\varepsilon,0,-\sqrt2\,d)$ and $(0,2\delta,\sqrt2\,b)$, with normal vector $(2\sqrt2\,d\delta,\,-2\sqrt2\,\varepsilon b,\,4\varepsilon\delta)$. For independent noises, $Q=\diag(q_1,q_2)$, this gives
\[
d_F(Q,\mathcal Q_A)=\frac{|\,\delta d\,q_1-\varepsilon b\,q_2\,|}{\sqrt{\delta^2d^2+\varepsilon^2b^2+2\varepsilon^2\delta^2}},
\qquad
\Big(\Cov(x,\log y)^2+\Cov(y,\log x)^2\Big)^{1/2}\ge\frac{d_F(Q,\mathcal Q_A)}{2\|A\|_{\mathrm{op}}} .
\]
The numerator is the same combination $\delta dq_1-\varepsilon bq_2$ that appears in the linear-noise covariance of Corollary~\ref{cor:pp-corr} below, and it vanishes exactly under the compatibility condition $q_1d/\varepsilon=q_2b/\delta$.
\end{example}

\subsection{The linear-noise approximation}\label{sec:linear}

To leading order in the noise intensity the correlations can be computed explicitly. Let $X^*=\diag(x^*)$ and $\mathcal M=AX^*$, the Jacobian of the drift of $u$ at $u^*=\log x^*$. With $H=CX^*$ we have $H\mathcal M+\mathcal M^\top H=X^*(CA+A^\top C)X^*<0$, so $\mathcal M$ is Hurwitz. In the linear-noise approximation \cite{vK07}, obtained by replacing $Q$ by $\epsilon^2Q$ and linearizing around $u^*$, the fluctuations $(u-u^*)/\epsilon$ are approximated by a stationary Ornstein--Uhlenbeck process with drift matrix $\mathcal M$, whose covariance $S$ is the unique solution of the Lyapunov equation
\begin{equation}\label{eq:lyap}
\mathcal MS+S\mathcal M^\top+Q=0 .
\end{equation}
Since $x\approx x^*(1+u-u^*)$ to first order, $K\approx\epsilon^2X^*S$, and \eqref{eq:lyap} is the linearization of \eqref{eq:exact-lyap}.

\begin{theorem}\label{thm:linear}
The solution $S$ of \eqref{eq:lyap} is diagonal if and only if
\[
Q=-\big(A\Delta+\Delta A^\top\big)
\]
for some positive diagonal matrix $\Delta$, that is, if and only if $Q$ is the compatible covariance $-k(AC'^{-1}+C'^{-1}A^\top)$ for the admissible matrix $C'=k\Delta^{-1}$ and some (any) $k>0$. In that case $S=\Delta X^{*-1}=k\,(C'X^*)^{-1}$.
\end{theorem}

\begin{proof}
If $Q=-(A\Delta+\Delta A^\top)$ with $\Delta$ positive diagonal, then $S=\Delta X^{*-1}$ is diagonal and $\mathcal MS+S\mathcal M^\top=A\Delta+\Delta A^\top=-Q$, since diagonal matrices commute; by uniqueness it is the solution. Conversely, if $S$ is diagonal, then it is positive definite, as the covariance of a process with nondegenerate noise, so $\Delta=SX^*$ is positive diagonal and $Q=-(\mathcal MS+S\mathcal M^\top)=-(A\Delta+\Delta A^\top)$. Admissibility of $C'$ follows as in the proof of Theorem~\ref{thm:exact}.
\end{proof}

Theorem~\ref{thm:linear} is the small-noise shadow of Theorem~\ref{thm:exact}. Note also that the diagonal form of $S$ is consistent with the exact result of Theorem~\ref{thm:stochastic}, since the variance of the logarithm of a $\mathrm{Gamma}(\alpha,\beta)$ variable is $\psi'(\alpha)=1/\alpha+O(\alpha^{-2})$.

\begin{corollary}\label{cor:pp-corr}
For the predator--prey system \eqref{eq:predator} with independent noises of intensities $q_1=\sigma_1^2$ and $q_2=\sigma_2^2$ (Stratonovich, with equilibrium $(x^*,y^*)$), the linear-noise covariance between $\log x$ and $\log y$ is
\[
S_{12}=\frac{\delta d\,q_1-\varepsilon b\,q_2}{2\,(\varepsilon x^*+\delta y^*)(\varepsilon\delta+bd)} .
\]
It vanishes if and only if $q_1d/\varepsilon=q_2b/\delta$, which is the compatibility condition of Section~\ref{sec:noise}, and it is positive if and only if $q_1d/\varepsilon>q_2b/\delta$, i.e.\ when the prey is noisier, relative to its self-limitation, than the predator.
\end{corollary}

\begin{proof}
Here $\mathcal M=\begin{psmallmatrix}-\varepsilon x^*&-by^*\\ dx^*&-\delta y^*\end{psmallmatrix}$ and $Q=\diag(q_1,q_2)$. The three scalar equations of \eqref{eq:lyap} are $2(\mathcal M_{11}S_{11}+\mathcal M_{12}S_{12})+q_1=0$, $2(\mathcal M_{21}S_{12}+\mathcal M_{22}S_{22})+q_2=0$ and $(\mathcal M_{11}+\mathcal M_{22})S_{12}+\mathcal M_{12}S_{22}+\mathcal M_{21}S_{11}=0$. Eliminating $S_{11}$ and $S_{22}$ gives
\[
S_{12}=\frac{\mathcal M_{11}\mathcal M_{12}\,q_2+\mathcal M_{22}\mathcal M_{21}\,q_1}{2\tr\mathcal M\,\det\mathcal M},
\]
and substituting $\tr\mathcal M=-(\varepsilon x^*+\delta y^*)$ and $\det\mathcal M=x^*y^*(\varepsilon\delta+bd)$ gives the formula.
\end{proof}

\section{Numerical illustration}\label{sec:numerics}

In this section we illustrate the results of Sections~\ref{sec:noise}--\ref{sec:incompatible} numerically. We first consider the predator--prey system \eqref{eq:predator} with compatible noise, and compare it with its uncoupled twin (Sections~\ref{sec:blind} and~\ref{sec:time}); we then compute its stationary entropy production as a function of self-limitation. Next we consider a three-species community with correlated compatible noise, and the coexistence threshold for It\^o noise. Finally we break compatibility in a controlled way (Section~\ref{sec:incompatible}).

\subsection{Numerical method}

In all cases we simulate the Stratonovich equations in the variables $u=\log x$,
\[
du=(r+Ae^{u})\,dt+\Lambda\,dW_t ,
\]
where $e^u$ is taken componentwise and $\Lambda$ is the Cholesky factor of the noise covariance $Q$; for It\^o noise, $r$ is replaced by $\rho-\frac12\diag(Q)$ (Remark~\ref{rem:moments-ito}). In these variables the noise is additive, so that the It\^o and Stratonovich forms of the equation coincide, and the positivity of the populations is automatic. We use the Euler--Maruyama scheme with time step $\Delta t=5\cdot10^{-3}$. All trajectories start at the equilibrium and are run for a burn-in time of $100$ before being recorded. Lagged covariances are computed by averaging over time and over independent trajectories, and $\dot R(0^+)$ is estimated entrywise by a quadratic least-squares fit of $R(\tau)$ on the eleven lags $\tau\in\{0,0.05,\dots,0.5\}$.

\subsection{Predator--prey with compatible noise: the stationary law}

We take $a=b=d=1$ and $c=\varepsilon=\delta=\frac12$, so that
\[
A=\begin{pmatrix}-\tfrac12&-1\\ 1&-\tfrac12\end{pmatrix},\qquad r=\begin{pmatrix}1\\ -\tfrac12\end{pmatrix},\qquad (x^*,y^*)=(0.8,0.6).
\]
The weights of Section~\ref{sec:predator} are $C=\diag(1,1)$, so that $AC^{-1}=A$, the Poisson part is $J=\begin{psmallmatrix}0&-1\\ 1&0\end{psmallmatrix}$ and the metric part is $\Gamma=\diag(\frac12,\frac12)$. The Jacobian of the drift of $u$ at equilibrium, $\mathcal M=AX^*=\begin{psmallmatrix}-0.4&-0.6\\ 0.8&-0.3\end{psmallmatrix}$, has eigenvalues $-0.35\pm0.69\,i$, so that fluctuations relax on a time scale of about $3$ and oscillate with a period of about $9$.

With $k=0.05$, the compatible noise of Corollary~\ref{cor:FDT} is $\Sigma=2k\Gamma=0.05\,I$, i.e.\ independent noises of intensity $\sigma_1^2=\sigma_2^2=0.05$, and $D=C/k=20\,I$. By Theorem~\ref{thm:stochastic}, the stationary law is
\[
x\sim\mathrm{Gamma}(16,20),\qquad y\sim\mathrm{Gamma}(12,20),
\]
independent, with means $0.8$ and $0.6$ and standard deviations $0.2$ and about $0.17$. By Corollary~\ref{cor:twin}, the uncoupled twin is obtained by setting the Poisson part to zero: it has $A'=-\frac12I$, $r'=-A'x^*=(0.4,0.3)$ and the same noise, i.e.\ it consists of the two independent logistic equations $dx=x(0.4-0.5x)\,dt+\sigma_1x\circ dW^1_t$ and $dy=y(0.3-0.5y)\,dt+\sigma_2y\circ dW^2_t$. By Theorem~\ref{thm:blind}, both systems have exactly the same stationary law.

\begin{figure}[t]
\centering
\includegraphics[width=\textwidth]{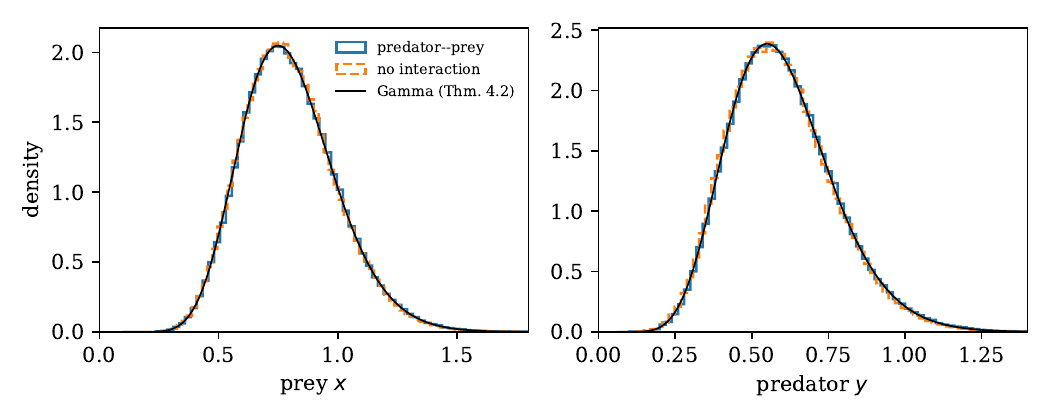}
\caption{Stationary marginals of the predator--prey system with compatible noise (solid) and of its uncoupled twin (dashed), with the Gamma densities of Theorem~\ref{thm:stochastic} (black). The two systems have the same noise and indistinguishable stationary laws.}
\label{fig:marginals}
\end{figure}

For each system we simulate $200$ independent trajectories, recorded for a time of $2000$ and sampled every $0.05$ time units, which gives $8\cdot10^6$ samples of the stationary law. Figure~\ref{fig:marginals} shows the histograms of the populations. The histograms of the interacting system and of its uncoupled twin are indistinguishable from each other and from the Gamma densities of Theorem~\ref{thm:stochastic}. The sample means are $(0.800,0.600)$ for the predator--prey system and $(0.799,0.599)$ for the twin, in agreement with Theorem~\ref{thm:moments}(a), and the sample correlations between $x$ and $y$ are $-0.0008$ and $0.0017$, respectively, in agreement with the independence of the populations. The predator--prey interaction, which is of order one, leaves no detectable trace in the stationary statistics.

\subsection{Predator--prey with compatible noise: the interaction in time}

\begin{figure}[t]
\centering
\includegraphics[width=\textwidth]{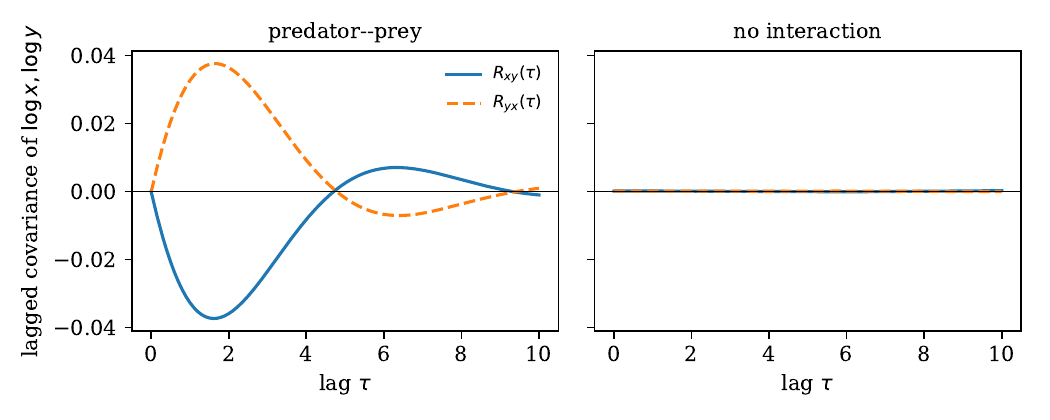}
\caption{Lagged cross-covariances $R_{xy}(\tau)=\Cov(\log x(t+\tau),\log y(t))$ and $R_{yx}(\tau)=\Cov(\log y(t+\tau),\log x(t))$ for the predator--prey system (left) and its uncoupled twin (right). The stationary laws are the same (Figure~\ref{fig:marginals}), but only the interacting system has time-asymmetric correlations.}
\label{fig:lagcov}
\end{figure}

Figure~\ref{fig:lagcov} shows the two lagged cross-covariances $R_{xy}(\tau)=\Cov(\log x(t+\tau),\log y(t))$ and $R_{yx}(\tau)=\Cov(\log y(t+\tau),\log x(t))$, computed from the same trajectories for lags between $0$ and $10$. For the uncoupled twin they vanish identically, as they must for independent populations. For the predator--prey system they vanish at $\tau=0$, but are of opposite signs at short lags: an excess of predators is followed by a deficit of prey ($R_{xy}<0$), and an excess of prey is followed by an excess of predators ($R_{yx}>0$). Both cross-covariances then oscillate and decay, with the period and relaxation time of the linearized dynamics. This time asymmetry is the signature of the Poisson part (Corollary~\ref{cor:identify}).

To recover the interaction matrix we apply Corollary~\ref{cor:identify}, estimating $K=\Cov(x,\log x)$ directly from the samples and computing $\dot R(0^+)K^{-1}$. The results are given in Table~\ref{tab:A}. For both systems the interaction matrix is recovered with errors of a few percent: the off-diagonal entries of the predator--prey system, which encode the interaction, are recovered to within $0.6\%$, and those of the twin are of order $10^{-3}$. The compatible version of the estimator, $\dot R(0^+)D$ with $D=\diag(\mathbb E[x_i]/\Var(x_i))$ estimated from the one-time moments, agrees with these values to within $0.002$. The other two predictions of Corollary~\ref{cor:identify} are also confirmed: the antisymmetric part of $\dot R(0^+)$ gives $\Omega_{12}=-0.0502$, against the exact value $-k=-0.05$, and $-(\dot R(0^+)+\dot R(0^+)^\top)$ gives the noise intensities $0.0514$ and $0.0518$, against the exact value $0.05$. The remaining discrepancies are consistent with the discretization error of the Euler--Maruyama scheme and with the finite length of the time series.

\begin{table}[t]
\centering
\begin{tabular}{lcc}
\toprule
 & exact $A$ & estimated $\dot R(0^+)K^{-1}$\\
\midrule
predator--prey & $\begin{pmatrix}-0.5&-1\\ 1&-0.5\end{pmatrix}$ & $\begin{pmatrix}-0.513&-1.006\\ 1.000&-0.517\end{pmatrix}$\\[3mm]
uncoupled twin & $\begin{pmatrix}-0.5&0\\ 0&-0.5\end{pmatrix}$ & $\begin{pmatrix}-0.497&0.002\\ 0.000&-0.498\end{pmatrix}$\\
\bottomrule
\end{tabular}
\vspace{2mm}
\caption{Interaction matrices recovered from the simulated time series by Corollary~\ref{cor:identify}. The estimate $\dot R(0^+)D$, with $D$ estimated from the ratios $\mathbb E[x_i]/\Var(x_i)$, agrees with these values to within $0.002$.}
\label{tab:A}
\end{table}

\subsection{Entropy production}

\begin{figure}[t]
\centering
\includegraphics[width=0.6\textwidth]{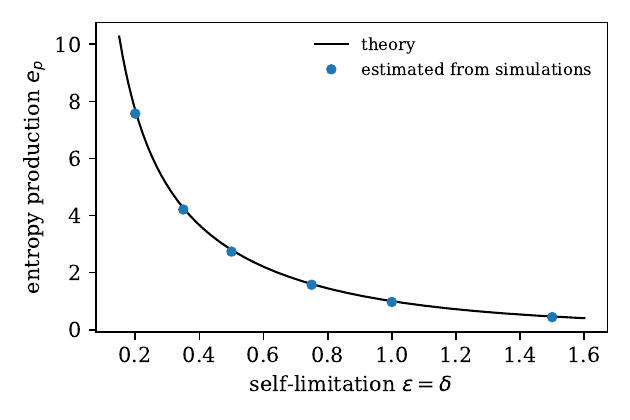}
\caption{Stationary entropy production rate of the predator--prey system with compatible noise as a function of self-limitation $\varepsilon=\delta$. Line: Theorem~\ref{thm:ep}. Points: estimates obtained from the simulated time series alone.}
\label{fig:ep}
\end{figure}

By Theorem~\ref{thm:ep}, the stationary entropy production rate of the predator--prey system with compatible noise is $e_p=bd(x^*/\delta+y^*/\varepsilon)$, which equals $2.8$ for the parameters above and vanishes for the uncoupled twin. To test the formula over a range of parameters, we keep $a=b=d=1$, $c=\frac12$ and $k=0.05$, and vary the self-limitation $\varepsilon=\delta$ over the six values $\{0.2,0.35,0.5,0.75,1,1.5\}$; the equilibrium $(x^*,y^*)=\big((\varepsilon+\frac12)/(1+\varepsilon^2),\,(1-\frac12\varepsilon)/(1+\varepsilon^2)\big)$ is interior for $\varepsilon<2$, and the compatible noise is $\Sigma=0.1\,\varepsilon\,I$. For each value we simulate $100$ trajectories recorded for a time of $1000$, and we estimate $e_p$ from the time series alone, by inserting in the first expression of Theorem~\ref{thm:ep} the estimates $\widehat\Omega=\frac12(\dot R(0^+)-\dot R(0^+)^\top)$, $\widehat\Sigma=-(\dot R(0^+)+\dot R(0^+)^\top)$ and $\widehat\alpha_i=\mathbb E[x_i]^2/\Var(x_i)$. Figure~\ref{fig:ep} compares these estimates with the theoretical values, which decrease from $7.69$ at $\varepsilon=0.2$ to $0.46$ at $\varepsilon=1.5$. The estimates follow the theoretical curve over the whole range, with a systematic underestimation between $1\%$ and $5\%$, of the same origin as the small bias in the estimate of $\dot R(0^+)$. In agreement with Theorem~\ref{thm:ep}, the entropy production grows without bound as self-limitation decreases, i.e.\ as the system approaches the conservative LV system.

\subsection{Three species with correlated compatible noise}

The two-species example has independent noises. To illustrate the general case of Theorem~\ref{thm:stochastic}, in which the compatible noise is correlated between species but the populations are still independent, and to test the identification of a larger interaction matrix, we consider a three-species community built with the parametrization of Theorem~\ref{thm:blind}(c). We take
\[
m=(0.8,0.6,0.5),\qquad D=\diag(20,25,30),
\]
\[
\Sigma=0.05\begin{pmatrix}1&0.6&0.3\\ 0.6&1&0.4\\ 0.3&0.4&1\end{pmatrix},\qquad
\Omega=0.05\begin{pmatrix}0&-1&0.5\\ 1&0&-0.8\\ -0.5&0.8&0\end{pmatrix},
\]
and $A=(\Omega-\frac12\Sigma)D$, $r=-Am$, which gives
\[
A=\begin{pmatrix}-0.5&-1.625&0.525\\ 0.7&-0.625&-1.5\\ -0.65&0.75&-0.75\end{pmatrix}.
\]
The signs of the off-diagonal entries describe a cycle of predation: species $2$ preys on species $1$ ($A_{12}<0<A_{21}$), species $1$ preys on species $3$ ($A_{31}<0<A_{13}$), and species $3$ preys on species $2$ ($A_{23}<0<A_{32}$). The noise covariance $\Sigma$ is compatible, with weights $C=kD$ for any $k>0$, and correlated, with correlation coefficients $0.6$, $0.3$ and $0.4$ between the noises acting on the pairs $(1,2)$, $(1,3)$ and $(2,3)$. Since $n=3$ is odd, the Poisson part has a Casimir function, and the stationary law depends on the choice of reference measure made in Remark~\ref{rem:measure}; with that choice, Theorems~\ref{thm:stochastic} and~\ref{thm:blind} predict the stationary law
\[
x_1\sim\mathrm{Gamma}(16,20),\qquad x_2\sim\mathrm{Gamma}(15,25),\qquad x_3\sim\mathrm{Gamma}(15,30),
\]
independent, in spite of the correlated noise. We compare this community with the system obtained by setting $\Omega=0$, which has the same noise and, by Theorem~\ref{thm:blind}, the same stationary law, but symmetric interactions:
\[
A'=-\tfrac12\Sigma D=\begin{pmatrix}-0.5&-0.375&-0.225\\ -0.3&-0.625&-0.3\\ -0.15&-0.25&-0.75\end{pmatrix},
\]
all of whose entries are negative, i.e.\ a community of competitors. This comparison differs from that of the previous subsections: the twin is not uncoupled, since the noise is not diagonal, but it has no Poisson part.

\begin{figure}[t]
\centering
\includegraphics[width=\textwidth]{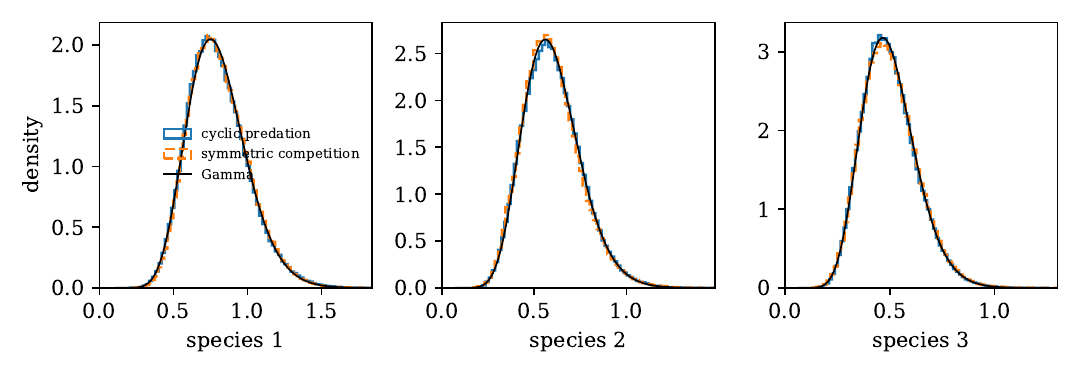}
\caption{Stationary marginals of the three-species community with cyclic predation (solid) and of the competitive community with the same noise and no Poisson part (dashed), with the Gamma densities of Theorem~\ref{thm:stochastic} (black). The noise is compatible and correlated between species.}
\label{fig:three}
\end{figure}

For each of the two communities we simulate $200$ trajectories recorded for a time of $1500$; the linearization $AX^*$ of the cyclic community has eigenvalues $-0.24\pm1.04\,i$ and $-0.67$. Figure~\ref{fig:three} shows the stationary marginals: those of the cyclic community and of the competitive one are indistinguishable from each other and from the predicted Gamma densities, and the sample means are $(0.800,0.600,0.500)$ and $(0.801,0.599,0.500)$. Table~\ref{tab:three} compares the correlations between the noises with those between the populations. Although the noises acting on different species are strongly correlated, the populations are uncorrelated to within $0.007$ in both communities, as predicted by Theorem~\ref{thm:stochastic}.

\begin{table}[t]
\centering
\begin{tabular}{lccc}
\toprule
pair of species & $(1,2)$ & $(1,3)$ & $(2,3)$\\
\midrule
correlation of the noises & $0.6$ & $0.3$ & $0.4$\\
correlation of the populations, cyclic predation & $-0.003$ & $-0.007$ & $-0.004$\\
correlation of the populations, competition & $-0.005$ & $0.001$ & $-0.004$\\
\midrule
$\Omega_{ij}$, exact & $-0.050$ & $0.025$ & $-0.040$\\
$\Omega_{ij}$, estimated (cyclic predation) & $-0.051$ & $0.026$ & $-0.041$\\
\bottomrule
\end{tabular}
\vspace{2mm}
\caption{Three-species communities with compatible, correlated noise. The populations are uncorrelated although the noises are not (Theorem~\ref{thm:stochastic}), and the reversible part $\Omega$ of the interaction, invisible in the stationary law, is recovered from the antisymmetric part of $\dot R(0^+)$ (Corollary~\ref{cor:identify}). For the competitive community the estimated $\Omega$ is zero to within $5\cdot10^{-5}$.}
\label{tab:three}
\end{table}

The time series, on the other hand, distinguish the two communities. Applying Corollary~\ref{cor:identify} to the cyclic community, we obtain
\[
\dot R(0^+)K^{-1}=\begin{pmatrix}-0.514&-1.631&0.561\\ 0.718&-0.648&-1.507\\ -0.648&0.782&-0.760\end{pmatrix},
\]
which reproduces $A$ with a maximal error of $0.036$, and recovers in particular the cyclic structure of the interaction. The estimated reversible part is given in Table~\ref{tab:three}, and the estimated noise covariance, $-(\dot R(0^+)+\dot R(0^+)^\top)$, reproduces the entries of $\Sigma$ to within $0.003$, including its off-diagonal entries. For the competitive community the same procedure reproduces $A'$ to within $0.006$, and its estimated reversible part vanishes.

\subsection{The coexistence threshold}

We now illustrate Proposition~\ref{prop:Ito} and the predator--prey example of Section~\ref{sec:noise} for It\^o noise. With the parameters of the first subsections, the It\^o noises $\sigma_1x\,dW^1_t$ and $\sigma_2y\,dW^2_t$ are compatible if and only if $\sigma_1^2=\sigma_2^2=k$, and condition \eqref{eq:coexistence} reduces to $(1-\frac k2)>\frac12(\frac12+\frac k2)$, i.e.\ $k<1$. We take $k\in\{0.5,0.9,1.2\}$ and simulate $20$ trajectories for each value up to time $600$. For $k=0.5$ and $k=0.9$, Proposition~\ref{prop:Ito}(ii) predicts a stationary distribution with means $(x^*_\sigma,y^*_\sigma)=(0.9,0.3)$ and $(0.98,0.06)$, respectively; the sample means are $(0.895,0.296)$ and $(0.994,0.060)$. For $k=1.2$, Proposition~\ref{prop:Ito}(iii) predicts that there is no stationary distribution in the open quadrant, and Remark~\ref{rem:invasion} predicts that the predator goes extinct at the rate given by its invasion rate, $-c-\frac{\sigma_2^2}2+d(a-\frac{\sigma_1^2}2)/\varepsilon=-0.3$. In the simulations, the slope of the mean of $\log y$ over $t\in[200,600]$ is $-0.294$, while the prey converges to the stationary law of the stochastic logistic equation, with sample mean $0.798$ against the predicted $(a-\frac{\sigma_1^2}2)/\varepsilon=0.8$.

\begin{figure}[t]
\centering
\includegraphics[width=\textwidth]{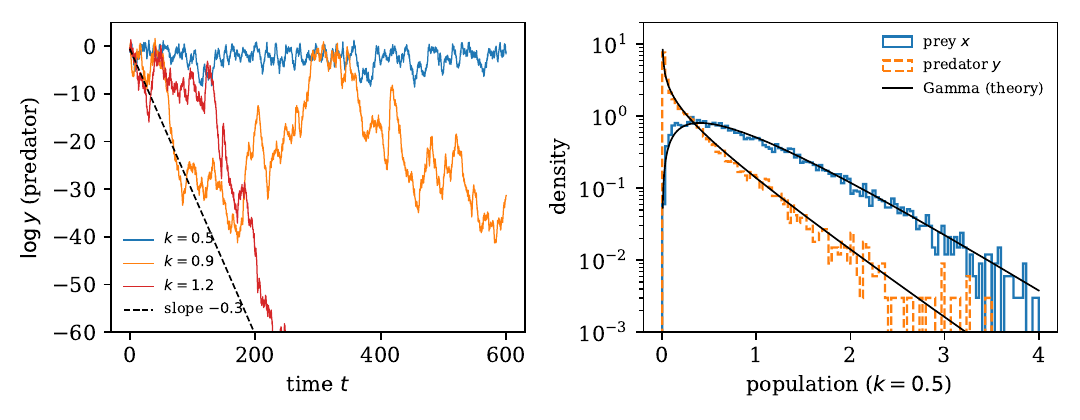}
\caption{Coexistence threshold for It\^o noise. Left: one trajectory of $\log y$ for $k=0.5$ and $k=0.9$, below the threshold $k=1$, and for $k=1.2$, above it; the dashed line has the slope $-0.3$ predicted by the invasion rate. Right: stationary densities for $k=0.5$ (logarithmic scale) and the predicted Gamma distributions.}
\label{fig:threshold}
\end{figure}

Figure~\ref{fig:threshold} shows one trajectory of $\log y$ for each value of $k$, and the stationary densities for $k=0.5$. Close to the threshold, for $k=0.9$, the predator persists but makes long excursions to very low abundances, in agreement with the small shape parameter $2\delta y^*_\sigma/\sigma_2^2\approx0.07$ of its stationary Gamma distribution. For $k=0.5$, the stationary densities agree with the Gamma distributions of the predator--prey example of Section~\ref{sec:noise}, with shape parameters $1.8$ and $0.6$ and rate $2$.

\subsection{Incompatible noise}

We finally break compatibility in a controlled way. In the predator--prey system with the parameters of the first subsections, we keep independent noises, $Q=\diag(q_1,q_2)$, fix the intensity of the noise acting on the predator at $q_2=0.05$, and vary the intensity $q_1$ of the noise acting on the prey. The departure from compatibility is measured by the ratio
\[
\kappa=\frac{q_1d/\varepsilon}{q_2b/\delta},
\]
which equals $1$ exactly in the compatible case of the first subsections; values $\kappa>1$ correspond to a prey that is noisier than its self-limitation allows, and values $\kappa<1$ to a predator in that situation. With our parameters $d/\varepsilon=b/\delta=2$, so that $\kappa=q_1/q_2$. We take the nine values $\kappa\in\{0.1,0.2,0.4,0.7,1,1.5,2.5,5,10\}$, which cover two decades, and for each of them we simulate $100$ trajectories, recorded for a time of $1000$ and sampled every $0.1$ time units. From these samples we compute the covariance matrix of $(\log x,\log y)$ and the matrix $K=\Cov(x,\log x)$.

\begin{figure}[t]
\centering
\includegraphics[width=0.6\textwidth]{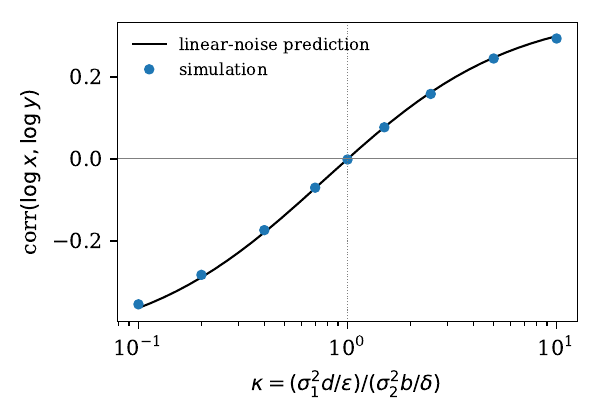}
\caption{Stationary correlation between $\log x$ and $\log y$ in the predator--prey system with independent noises, as a function of the incompatibility ratio $\kappa$. Points: simulations. Line: linear-noise approximation \eqref{eq:lyap}. The compatible case is $\kappa=1$.}
\label{fig:incompatible}
\end{figure}

\emph{Correlations in the linear-noise approximation.} Figure~\ref{fig:incompatible} shows the stationary correlation between $\log x$ and $\log y$ as a function of $\kappa$, together with the prediction of the linear-noise approximation, obtained by solving the Lyapunov equation \eqref{eq:lyap} numerically. The correlation vanishes at $\kappa=1$, as required by Theorem~\ref{thm:linear}; it is negative for $\kappa<1$ and positive for $\kappa>1$, as predicted by Corollary~\ref{cor:pp-corr}; and it reaches about $-0.36$ at $\kappa=0.1$ and $0.3$ at $\kappa=10$. Since the interaction matrix is the same for all values of $\kappa$, these correlations are produced by the noise alone. The simulated values agree with the linear-noise prediction to within $0.01$ over the whole range.

\begin{figure}[t]
\centering
\includegraphics[width=0.6\textwidth]{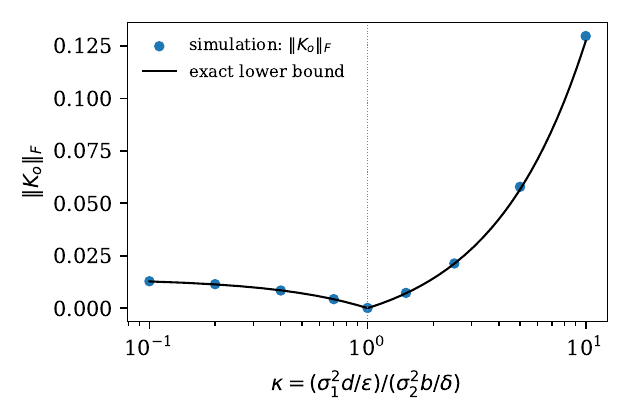}
\caption{Off-diagonal part $\|K_o\|_F$ of $K=\Cov(x,\log x)$ in the predator--prey system with independent noises, as a function of the incompatibility ratio $\kappa$. Points: simulations. Line: the lower bound of Proposition~\ref{prop:bound}, which vanishes only in the compatible case $\kappa=1$.}
\label{fig:bound}
\end{figure}

\emph{The exact identity and the lower bound.} The exact identity \eqref{eq:exact-lyap} holds for every noise intensity, without linearization. In the simulations, the relative residual $\|AK+K^\top A^\top+Q\|_F/\|Q\|_F$ is below $0.016$ for all values of $\kappa$, which is the level of the sampling and discretization errors. For our parameters $\|A\|_{\mathrm{op}}=\sqrt5/2$, and the formula of Example~\ref{ex:pp-bound} gives
\[
d_F(Q,\mathcal Q_A)=\frac{|q_1-q_2|}{2\sqrt{0.625}},\qquad
\frac{d_F(Q,\mathcal Q_A)}{2\|A\|_{\mathrm{op}}}\approx0.0141\,|\kappa-1| ,
\]
so that the lower bound of Proposition~\ref{prop:bound} is a V-shaped function of $\kappa$ that vanishes only in the compatible case. Figure~\ref{fig:bound} compares it with the simulated values of $\|K_o\|_F=(\Cov(x,\log y)^2+\Cov(y,\log x)^2)^{1/2}$. At $\kappa=1$ we obtain $\|K_o\|_F=7\cdot10^{-5}$, in agreement with Corollary~\ref{cor:independence}. Away from $\kappa=1$, the simulated values exceed the bound by at most $3\%$, and at one value of $\kappa$ fall below it by $1\%$, which is within the numerical error. In this example the lower bound is therefore nearly sharp over two decades of $\kappa$.

The near sharpness of the bound is not a general feature. For the three-species cyclic community of the previous subsection with the incompatible noise $Q=0.05\,I$, obtained by suppressing the correlations of the compatible noise, the exact identity holds with a relative residual of $0.015$, the populations become correlated, with correlation coefficients $-0.22$, $-0.27$ and $-0.30$ for the pairs $(1,2)$, $(1,3)$ and $(2,3)$, and $\|K_o\|_F=0.035$, whereas the lower bound, computed by solving the least-squares problem for $d_F(Q,\mathcal Q_A)$ numerically, is $0.011$. The bound is satisfied, but it is smaller than the actual value by a factor of about $3$.
\section{Concluding remarks}

We have studied Lotka--Volterra systems with self-limitation from the point of view of metriplectic geometry, and we have shown that this point of view clarifies what fluctuating abundance data can and cannot reveal about the interactions between species. Our results can be summarized as follows.

First, dissipative LV systems are metriplectic exactly when Volterra's condition $CA+A^\top C\le0$ holds, with a log-canonical Poisson bivector built from the skew-symmetric part of $AC^{-1}$ and a metric built from its symmetric part; no other choice of bivector or metric can avoid this condition. Second, the fluctuation--dissipation relation of stochastic GENERIC selects a family of compatible environmental noises, whose intensities are proportional to self-limitation and for which the energy is conserved exactly. Third, for these noises the stationary law is an explicit Gibbs measure, a product of Gamma distributions, for a class of nonsymmetric interactions that includes predator--prey interactions, even when the noises acting on different species are correlated; in the statistical physics of disordered ecosystems such measures were available only for symmetric interactions.

These three results lead to a dictionary between geometry and observation. The Volterra function determines the stationary law, the metric determines the noise, and the Poisson bivector is invisible at stationarity: every interaction matrix in an open cone of dimension $n^2$ produces the same stationary law, which, for independent noises, is also the law of a system of uncoupled stochastic logistic equations. The Poisson bivector is, however, visible in time. The exact identity $\dot R(0^+)=A\Cov(x,\log x)$, valid for any noise, recovers the interaction matrix from one-time moments and short-lag correlations of the log-abundances, and the reversible part of the interaction is exactly the time asymmetry of these correlations. Correspondingly, the stationary entropy production rate is carried entirely by the Poisson part, whereas in the deterministic system the entropy is produced by the metric part. Finally, for certain noise, the exact identity $AK+K^\top A^\top+Q=0$ shows that the stationary populations are independent if and only if the noise is compatible, and bounds the correlations between species from below by the distance of the noise to the compatible family.

For the analysis of ecological data, these results have both
a negative and a positive side. On the negative side,
stationary abundance statistics, such as Gamma-distributed
and uncorrelated fluctuations, may conceal interspecific
interactions. Under independent environmental noise,
this ambiguity concerns interactions of predator--prey
type, whereas other interaction structures may leave
detectable signatures in stationary correlations.
On the positive side, short-lag correlations of
log-abundances identify the interaction matrix exactly,
without the linearization on which multivariate
autoregressive methods rely. Moreover, stationary
correlations between species reflect the interplay
between noise and dissipation, rather than providing
a direct measure of interaction strength.

The scope of these conclusions should be stated clearly. Compatible noise is an exactly solvable reference case, not a claim about real environmental noise: for a single species every environmental noise is compatible, but for $n\ge2$ compatibility ties the noise intensities to self-limitation, and generic noise does not satisfy it (Remark~\ref{rem:family}). Two of our results do not depend on compatibility: the moment identity of Theorem~\ref{thm:moments} and the exact identity of Theorem~\ref{thm:exact} hold for any constant noise covariance, and the latter turns compatibility into a testable hypothesis through Corollary~\ref{cor:independence}. Our results also concern environmental noise; whether a comparable class of nonsymmetric interactions with an explicit stationary law exists for demographic noise is open (Remark~\ref{rem:disordered}).

Several further questions remain. The estimator of Corollary~\ref{cor:identify} should be analysed statistically for sampled, finite and noisy time series, and tested on microbial time series, where it would complement inference methods based on the stochastic logistic model. It would be interesting to obtain an upper bound complementing Proposition~\ref{prop:bound}, to understand when the lower bound is nearly sharp, as in our two-species example, and when it is not, as in our three-species example, and to describe how the product structure of the stationary measure is deformed for nearly compatible noise, perturbatively in $d_F(Q,\mathcal Q_A)$; the persistence and extinction thresholds for such noise are covered by \cite{HN18}, as Remark~\ref{rem:invasion} illustrates. The case of a positive semidefinite but singular $\Gamma$, which includes species without self-limitation, requires a separate analysis of uniqueness. On the geometric side, the decomposition of the dynamics into a skew-symmetric and a symmetric part is also the starting point of a port-Hamiltonian description, in which harvesting or immigration act as inputs, and it would be natural to study controls that shape the stationary law or, by acting only on the reversible part, accelerate the convergence to it without changing it. Finally, the coordinates $u=\log x$, in which both $\Pi$ and $\Gamma$ are constant, should be the natural setting for stochastic integrators that preserve the energy, the entropy production and the positivity of the populations exactly.



\begin{thebibliography}{99}

\bibitem{BBC18}
G.~Biroli, G.~Bunin, C.~Cammarota,
Marginally stable equilibria in critical ecosystems,
\emph{New J. Phys.} \textbf{20} (2018) 083051.

\bibitem{ARCB21}
A.~Altieri, F.~Roy, C.~Cammarota, G.~Biroli,
Properties of equilibria and glassy phases of the random Lotka--Volterra model with demographic noise,
\emph{Phys. Rev. Lett.} \textbf{126} (2021) 258301.

\bibitem{Bun17}
G.~Bunin,
Ecological communities with Lotka--Volterra dynamics,
\emph{Phys. Rev. E} \textbf{95} (2017) 042414.

\bibitem{DFO98}
P.~Duarte, R.L.~Fernandes, W.M.~Oliva,
Dynamics of the attractor in the Lotka--Volterra equations,
\emph{J. Differential Equations} \textbf{149} (1998) 143--189.

\bibitem{GMM71}
N.S.~Goel, S.C.~Maitra, E.W.~Montroll,
On the Volterra and other nonlinear models of interacting populations,
\emph{Rev. Mod. Phys.} \textbf{43} (1971) 231--276.

\bibitem{Gri20}
J.~Grilli,
Macroecological laws describe variation and diversity in microbial communities,
\emph{Nat. Commun.} \textbf{11} (2020) 4743.

\bibitem{GH25}
M.Y.~Gueddari, W.~Hachem,
Spin glass analysis of the invariant distribution of a Lotka--Volterra SDE with a large random interaction matrix,
arXiv:2510.15754 (2025).

\bibitem{GO97}
M.~Grmela, H.C.~\"Ottinger,
Dynamics and thermodynamics of complex fluids. I. Development of a general formalism,
\emph{Phys. Rev. E} \textbf{56} (1997) 6620--6632.

\bibitem{Guh07}
P.~Guha,
Metriplectic structure, Leibniz dynamics and dissipative systems,
\emph{J. Math. Anal. Appl.} \textbf{326} (2007) 121--136.

\bibitem{HN18}
A.~Hening, D.H.~Nguyen,
Coexistence and extinction for stochastic Kolmogorov systems,
\emph{Ann. Appl. Probab.} \textbf{28} (2018) 1893--1942.

\bibitem{HS98}
J.~Hofbauer, K.~Sigmund,
\emph{Evolutionary Games and Population Dynamics},
Cambridge University Press, 1998.

\bibitem{HHS93}
C.-R.~Hwang, S.-Y.~Hwang-Ma, S.-J.~Sheu,
Accelerating Gaussian diffusions,
\emph{Ann. Appl. Probab.} \textbf{3} (1993) 897--913.

\bibitem{IIP11}
G.~Ivan, M.~Ivan, C.~Pop,
Numerical integration and synchronization for the 3-dimensional metriplectic Volterra system,
\emph{Math. Probl. Eng.} \textbf{2011} (2011) 723629.

\bibitem{Ives03}
A.R.~Ives, B.~Dennis, K.L.~Cottingham, S.R.~Carpenter,
Estimating community stability and ecological interactions from time-series data,
\emph{Ecol. Monogr.} \textbf{73} (2003) 301--330.

\bibitem{JQQ04}
D.-Q.~Jiang, M.~Qian, M.-P.~Qian,
\emph{Mathematical Theory of Nonequilibrium Steady States},
Lecture Notes in Mathematics 1833, Springer, 2004.

\bibitem{Kau84}
A.N.~Kaufman,
Dissipative Hamiltonian systems: a unifying principle,
\emph{Phys. Lett. A} \textbf{100} (1984) 419--422.

\bibitem{Ker57}
E.H.~Kerner,
A statistical mechanics of interacting biological species,
\emph{Bull. Math. Biophys.} \textbf{19} (1957) 121--146.

\bibitem{Ker59}
E.H.~Kerner,
Further considerations on the statistical mechanics of biological associations,
\emph{Bull. Math. Biophys.} \textbf{21} (1959) 217--255.

\bibitem{Kha12}
R.~Khasminskii,
\emph{Stochastic Stability of Differential Equations}, 2nd ed., with contributions by G.N.~Milstein and M.B.~Nevelson,
Stochastic Modelling and Applied Probability~66, Springer, 2012.

\bibitem{MSR03}
X.~Mao, S.~Sabanis, E.~Renshaw,
Asymptotic behaviour of the stochastic Lotka--Volterra model,
\emph{J. Math. Anal. Appl.} \textbf{287} (2003) 141--156.

\bibitem{Mor84}
P.J.~Morrison,
Bracket formulation for irreversible classical fields,
\emph{Phys. Lett. A} \textbf{100} (1984) 423--427.

\bibitem{Mor86}
P.J.~Morrison,
A paradigm for joined Hamiltonian and dissipative systems,
\emph{Physica D} \textbf{18} (1986) 410--419.

\bibitem{Ott05}
H.C.~\"Ottinger,
\emph{Beyond Equilibrium Thermodynamics},
Wiley, 2005.

\bibitem{PS25}
M.A.~Peletier, M.~Seri,
A geometric formulation of GENERIC stochastic differential equations,
arXiv:2509.09566 (2025).

\bibitem{Pin95}
R.G.~Pinsky,
\emph{Positive Harmonic Functions and Diffusion},
Cambridge Studies in Advanced Mathematics~45, Cambridge University Press, 1995.

\bibitem{Pla95}
M.~Plank,
Hamiltonian structures for the $n$-dimensional Lotka--Volterra equations,
\emph{J. Math. Phys.} \textbf{36} (1995) 3520--3534.

\bibitem{RBS15}
L.~Rey-Bellet, K.~Spiliopoulos,
Irreversible Langevin samplers and variance reduction: a large deviations approach,
\emph{Nonlinearity} \textbf{28} (2015) 2081--2103.

\bibitem{Tak96}
Y.~Takeuchi,
\emph{Global Dynamical Properties of Lotka--Volterra Systems},
World Scientific, 1996.

\bibitem{vK07}
N.G.~van Kampen,
\emph{Stochastic Processes in Physics and Chemistry}, 3rd ed.,
Elsevier, 2007.

\bibitem{Vol31}
V.~Volterra,
\emph{Le\c{c}ons sur la th\'eorie math\'ematique de la lutte pour la vie},
Gauthier-Villars, Paris, 1931.

\end{thebibliography}
\end{document}